\documentclass[11pt,a4paper]{article}

\def\ptmiss{\ensuremath{\vec{p}_\mathrm{T}^{\mathrm{miss}}}}
\def\rpv{\mbox{${\not\!\!R_p}$}}

\usepackage{jheppub}
\usepackage{subfigure}
\usepackage{epstopdf}
\usepackage{slashed}
\usepackage{feynmp}

\author[a]{B.~C.~Allanach}
\affiliation[a]{Department of Applied Mathematics and Theoretical Physics, Centre for Mathematical Sciences, University of Cambridge, Wilberforce Road, Cambridge CB3 0WA, United Kingdom}
\emailAdd{B.C.Allanach@damtp.cam.ac.uk}
\author[b]{Ben Gripaios}
\affiliation[b]{Cavendish Laboratory, University of Cambridge, JJ Thomson
  Avenue, Cambridge CB3 0HE, United Kingdom}
\emailAdd{gripaios@hep.phy.cam.ac.uk}

\title{Hide and Seek With Natural Supersymmetry at the LHC} 

\abstract{Gluinos that result in classic large missing transverse momentum
  signatures at 
the LHC 
have been excluded by 2011 searches if they are lighter than around 800
GeV.
This adds to the tension between experiment and supersymmetric
solutions of the naturalness problem, since the gluino is required to be
light if the electroweak scale is to be natural.
Here, we examine natural scenarios where supersymmetry is present, but
was hidden from 2011 searches due to violation of $R$-parity and the
absence of a large missing transverse momentum signature. Naturalness suggests
that 
third generation states should dominate gluino decays and we argue
that this leads to a generic signature in the form of same-sign,
flavour-ambivalent leptons, without large missing transverse momentum. As a
result, searches in this channel are 
able to cover a broad range of scenarios with some generality and one should
seek gluinos that decay in this way
with masses below a TeV. We encourage the LHC experiments
to tailor a search for supersymmetry in this form. 
We consider a specific case that is good at hiding: baryon number violation, 
and estimate that the most constraining existing search from 2011 data implies a
lower bound on the gluino mass of 550 GeV. 
}

\begin{document}
\maketitle
\section{Introduction}
Softly-broken supersymmetry (SUSY) has long provided an elegant dogma for
natural electroweak symmetry breaking, providing a mechanism by which
the electroweak scale may be stabilised in the presence of radiative
corrections from new physics at arbitrarily large mass scales.
But even fervent disciples have seen their faith in SUSY shaken in
recent times by the lack of evidence therefor. 

Indeed, a generic, natural, supersymmetric theory predicts
superpartners with masses liberally sprinkled around the weak scale
(measured by, say, $m_Z$) and so
negative searches for direct production of superparticles at LEP (and
for the Higgs boson)
already put strong constraints on generic supersymmetric theories. 
In a given theory, these constraints can be avoided by pushing up
the mass scale of superparticles and carefully choosing the parameters
to obtain the observed value of $m_Z$, but this is in itself nothing other than
a fine tuning, corresponding to a residual 
``little hierarchy'' between $m_Z$ and the mass scale of
superpartners.\footnote{Alternatively, one might imagine that a
  baroque theory exists in which the measured fine tuning, according
  to some definition, turned out to be small. But then one would be
  left wondering why Nature would choose such a theory
  apparently solely for the purpose of hiding physics beyond the
  Standard Model (SM) from our current generation of experiments.}

The situation has worsened with the advent of the 7 TeV LHC, where a
suite of negative
searches \cite{Aad:2011ib,ATLAS:2011ad,Aad:2011qa,Aad:2011cw,Chatrchyan:2011zy} has pushed lower bounds on superpartner masses, in
certain scenarios, beyond a TeV; though fine tuning is inherently a
subjective notion, many of us no doubt feel
that a psychological Rubicon has been crossed when bounds on superpartner masses
reach the terascale. 

As such, now would seem to be a good point to re-assess our agenda for
future SUSY searches. Should searches be
abandoned? Should we continue with existing searches, ratcheting up
the luminosity and beam energy, and interpret the results in terms
of the same models? Or should we pluck new models
that purport to solve the hierarchy problem up to high scales out of the
swamp of possibilities and design new searches for these?
None of these strategies seems especially appealing. 

A different suggestion has recently been put forward in Ref.~\cite{Brust:2011tb}. There, it is argued that, since the LHC directly probes
energy scales well beyond the weak scale (up to a cut-off, $\Lambda$,
less than a few TeV or so),
we should try to discern whether there are any superparticles present
below $\Lambda$ that stabilise the dynamics up to $\Lambda$, without regard for what further dynamics might be required, beyond the
reach of the LHC, to solve the `big' hierarchy problem. This would seem to be a laudable goal from the point of view
of the legacy of the LHC: if such superparticles are found at the LHC,
then detailed investigation of them will become a priority for
the next generation of experiments; if such superparticles are not
found and if we can be reasonably confident that they are not there,
then this will be a strong indicator that supersymmetric naturalness
is not a good guide for
predicting what (if any) physics may lie within our future reach.

This goal of ruling out supersymmetric naturalness within the reach of
the LHC would, moreover, seem to be achievable, by and large, if we make the
assumption of ``minimality''
\cite{Brust:2011tb}: that only those degrees of freedom that are required by
naturalness are present in the low energy spectrum (with perhaps just
a few other, auxiliary, states). 
This assumption is motivated not only phenomenologically (after all,
we have yet to see {\em any}\/ substantial evidence for superparticles and indeed the many direct and
indirect constraints on TeV- scale dynamics are most easily alleviated
by positing as few new TeV-scale particles as possible), but also by
the existence of theoretical models whose dynamics is such that the lightest
superparticles are automatically those
that couple most strongly to the Higgs sector
\cite{Gherghetta:2003wm,Sundrum:2009gv,Redi:2010yv,Gherghetta:2011wc}. (For other models
giving rise to such dynamics, see
\cite{Dimopoulos:1995mi,Pomarol:1995xc,Barbieri:1995uv,Cohen:1996vb,Barbieri:2010pd,Craig:2011yk,Barbieri:2011ci}.)
The experimental implications of the minimality assumption are that we are left
with a relatively small number of possible final states on which to focus
LHC searches.

Happily, many of these final states are already being sought and much
effort is being put into interpreting those searches in terms of a
generic exclusion of supersymmetric models. But it is important that
searches cover {\em all}\/ feasible scenarios and that we strive to
obtain the best possible sensitivity in each case. 

Here, we wish to focus on a generic scenario that is not currently
under examination, as far as we are aware. The scenario involves light
gluinos, which are required by naturalness and in consequence have
significant production cross-section at the LHC\@. The scenario differs
from canonical ones in that $R$-parity conservation (which is not required by
naturalness) is not assumed, such that conventional searches based on large
missing transverse momentum ($|\ptmiss|$)
signatures fail. Nevertheless, there is sensitivity at the LHC,
because pairs of Majorana gluinos sometimes decay into
pairs of same-sign leptons (plus hadrons). The reasons that we arrive at
leptons are two-fold. Firstly, naturalness and flavour physics
constraints suggest that third
generation quarks and squarks are lightest and that their couplings
are largest, meaning that
top
quarks are often present in gluino decays. A significant fraction of
these further decay into isolated leptons.
Secondly, the $R$-parity violating couplings by which superpartners
ultimately decay frequently involve either leptons or (again by naturalness
and flavour arguments) top-quarks. So a search for same-sign leptons
(which should include $\tau$ leptons) should cover many of the
possible gluino decays. In fact we find only one, rather special, case in which a
significant fraction of same-sign leptons from gluino decays is not
automatic: for that to happen, the gluino must decay
into a left-handed bottom squark. This decay is unlikely to
overwhelmingly dominate, given that the gluino couples with equal
strength to left-handed top and bottom squarks and given that, at  
tree-level, the difference in the squared masses of these squarks is
only $m_{\tilde{t}_L}^2 - m_{\tilde{b}_L}^2 = M_W^2 \cos 2\beta$. Even
if the decay to a left-handed sbottom does predominate, 
the decay of the latter must proceed via a charged Higgs boson to a virtual right-handed top
squark, which in turn must decay via the superpotential operator $W \supset U_3D_iD_j$ into light
quark jets. Then, the only possible source of an isolated lepton comes
from the charged Higgs decay. This decay will not result in a source
of leptons only if the charged Higgs is below threshold to decay to $t\overline{b}$
and only if the CKM-suppressed decay to $c\overline{b}$ dominates over
the decay to $\tau \overline{\nu}_\tau$. This last condition requires
that
$\tan \beta \ll \frac{m_t}{m_\tau}V_{cb} \approx 3$, which seems
unlikely, at least in the context of the minimal supersymmetric SM
(MSSM), due to higgs constraints from LEP. 

The outline of the paper is as follows. In the next section, we review
the arguments for which superparticles should be light and catalogue
the possible LHC scenarios, showing that many are already being
covered. In \S\ref{sec:rpv}, we consider the case of
gluino pair-production followed by $R$-parity-violating decays in detail,
showing that most cases can be covered by a search for a pair of same-sign $e$,
$\mu$ or $\tau$ appearing in any flavour combination.
In \S\ref{sec:current}, we
argue that existing LHC searches for low-scale quantum gravity and
canonical, $R$-parity-conserving SUSY in final states involving same sign
leptons give the best current sensitivity and provide a rough estimate of the
resulting bounds on the gluino mass from three such searches, in a scenario where the
gluino decays via a top squark which subsequently decays via a
baryon-number-violating operator. We then argue, in
\S\ref{sec:future},  that
dedicated searches could provide significantly stronger bounds and we
encourage the LHC collaborations to carry them out. 
\section{Natural supersymmetry at the LHC}
Which states then, ought to be within reach of the LHC\@? The largest
contribution to the quadratic divergence in the Higgs mass parameter
comes from a loop of top quarks via the Yukawa coupling. To cancel
this using supersymmetric dynamics, we need both the right-handed top
squark, $\tilde{u}^3_R$, and the left-handed $SU(2)_L$ doublet
containing top and bottom squarks, $\tilde{Q}^3_R$, to be light. 

The next largest quadratic divergences come from one-loop diagrams
involving $W$-bosons and the Higgs itself, but already the couplings
involved are small enough that one can imagine them being cancelled by
superpartners that are beyond the reach of the LHC\@. This is especially
true when
one takes into account the fact that these are colour-singlet
states with commensurately small direct production cross-sections in $pp$
collisions.

The minimal dynamics, then, contains only left- and right-handed top
squarks, together with the left-handed bottom squark. Search
strategies for these depend on when and how they decay. If they are
sufficiently long-lived to reach the detector before decaying (perhaps
because of some symmetry, like $R$-parity, either approximate or
exact)\footnote{We do not concern ourselves here with potential
  cosmological issues arising from stable or long-lived coloured
  particles.}
then the exotic `$R$-hadrons' into which they are confined may be detected either by their interactions with the
detector or by their eventual decays \cite{Fairbairn:2006gg}. Existing
searches, though subject to large uncertainties due to our poor
understanding of the properties of $R$-hadrons, already exclude masses
up to around 500 GeV \cite{CMS2}, so there is hope that the TeV Rubicon
will eventually be crossed. On the other hand, stops and sbottoms may
decay promptly. For the lightest state to decay, $R$-parity must be
violated.
 We write the $R-$parity violating part of the 
renormalisable MSSM superpotential~\cite{Allanach:2009bv} (in the basis where
the superfields have been rotated such that the quarks contained within are in
the mass basis)
\begin{equation} 
W_{\rpv}= \frac{1}{2} \lambda_{ijk} L_i L_jE_k +
\lambda'_{ijk} L_i Q_j {D}  - \kappa^i L_i H_2+
\frac{1}{2} \lambda''_{ijk} {U}_i{D}_j{D}_k,
 \label{superpot1} 
\end{equation} 
where we have suppressed the gauge indices, $i,j,k$ are flavour indices, and 
$L_i$,$Q_i$,$U_i$,$D_i$,$E_i$ are chiral superfields containing left-handed
fermions: lepton doublets,
quark doublets,  anti up-quarks, anti down quarks and  positrons,
respectively. 
The $U_i D_j D_k$ operators break baryon number, whereas the others break
lepton number. We do not expect both sets of operators to be present
simultaneously, since 
proton decay would be predicted to be much faster than is observed.
However, several discrete gauge symmetries have been proposed which may ban 
one set of operators~\cite{Ibanez:1991pr,Allanach:2003eb} as alternatives to
$R-$parity. 
As has long been known \cite{Sakai:1981pk,Weinberg:1981wj}, and has
been re-stressed in \cite{Brust:2011tb}, $R$-parity itself does not forbid
dimension-five operators 
in the superpotential, such as $W \supset \frac{Q_iQ_jQ_kL_l}{\Lambda}$, by
which the proton may decay. 
So $R$-parity does not seem to be
well-motivated in the context of an effective theory in
which we profess ignorance of the dynamics beyond the reach of the
LHC, such that the cut-off $\Lambda$ is just a few TeV. 

If stops or sbottoms do
decay promptly, they will presumably do so via the $R$-parity
violating operators in the superpotential of lowest dimension, {\em viz.}\/
either via $\lambda^{\prime}_{3jk}$ or $\lambda^{\prime\prime}_{3mn}$. 
The
couplings  
$\lambda^{\prime}_{3jk}$ or $\lambda^{\prime\prime}_{3mn}$ endow the squarks
with 
leptoquark or diquark properties, respectively, and the
 final states of
interest then depend on the values of the remaining indices. 
In the case of leptoquark-like states, only couplings to the
third-generation are likely to be sizable if constraints coming from
flavour physics are to be satisfied \cite{Gripaios:2009dq}, leading to
final states which are pairwise combinations of $t\tau$, $t\nu_\tau$,
$b\tau$, and $b\nu_\tau$. The LHC
sensitivities for such leptoquarks are studied in
\cite{Gripaios:2010hv}, but are unlikely to rise beyond a few hundred
GeV. A search in the $2b2\nu_\tau$ final state was recently performed
in \cite{CMS4}.
In the case of diquark-like states, it was shown in
\cite{Giudice:2011ak} that for a right-handed top
squark (which couples antisymmetrically in flavour indices to a pair
of down quarks) any one of the three couplings
$\lambda^{\prime\prime}_{3mn}$ could be sizable. This opens up the
possibility of single production (leading to sensitivity in di-jet
resonance searches) or pair production followed by decay into
four jets (possibly with heavy-flavour), for which existing searches may be
found in \cite{CMS3,Aad:2011yh}.

We next ask what may happen if (a small number of) additional superparticles
are present. For the purposes of LHC phenomenology, it is easiest to
classify these by their colour charges. Colour singlets have low
production cross section and are most likely to affect
phenomenology through their appearance in decays of the produced top
and bottom squarks. Perhaps the most likely such particles are
Higgsinos, whose masses may arise from the $R-$parity conserving
superpotential term  
$W \supset \mu H_u H_d$. This term also contributes to the Higgs
scalar potential and thus is partly responsible for
setting the weak scale, via $\frac{m_Z^2}{2} \simeq -|\mu|^2 - m_{H_u}^2$, where
$m_{H_u}^2$ is a soft SUSY breaking mass term. Thus, unless $\mu$ is of order $m_Z$
(implying a Higgsino mass of order $m_Z$) an apparently unnatural fine tuning
between supersymmetric and SUSY-breaking terms is
required. However, this argument for the existence of a light
Higgsino cannot be considered watertight, even putting aside the subjectivity of naturalness
considerations:
in an effective theory with low cut-off $\Lambda$,
Higgsinos may also receive contributions to their masses from
higher-dimensional, SUSY- breaking terms, for example from the
K\"{a}hler potential term $K \supset
\frac{X^\dagger X}{\Lambda^3} D_\alpha H_u D^\alpha H_d$, where $X
\supset F \theta^2$ is a SUSY-breaking spurion.

If Higgsinos or other colour singlets are present, then we have
the possibility of top and bottom squark decays to these charged- or
electrically-neutral states, together with top and bottom
quarks. In the case that these are long-lived, we must look either for
$|\ptmiss|$ in final states or charged tracks in the detector; if they
decay promptly via $R$-parity-violating operators, we may expect
various combinations of jets, leptons and $|\ptmiss|$.

In all cases without charged tracks of new, heavy particles, it seems unlikely
that we will ever get sensitivity to squark masses much above several
hundred GeV, the reason being that the
production cross section for individual squarks is too low in the
region in which signals have a clean kinematic separation from
backgrounds. As an example, the cross section to pair-produce a top squark
with mass equal to the mass of the top quark is
reduced by a factor of around eight, compared to the
$t\overline{t}$  production cross section, due to spin and threshold
effects. The problem is compounded by the
fact that the
$|\ptmiss|$ final states, like $2\tilde{t}
\rightarrow 2(t + \chi^0)$, are kinematically very similar to
backgrounds from SM $t\overline{t}$ production. As a result, the current limits
on squark masses in such scenarios are extremely low, below $m_t$ even
 \cite{Aaltonen:2009sf,Kats:2011it,Papucci:2011wy}.\footnote{An exclusion of up to a
   few hundred GeV can be
   obtained if more than one squark is light
   \cite{Brust:2011tb,Papucci:2011wy}. Another possible loophole to this
argument is that, for rather heavy squarks,
one can hope to compensate for the reduced cross section by use of
jet substructure techniques applied to the boosted decay products of
the squarks.}

Adding more coloured particles may help to increase the LHC
cross-section. Adding colour triplets ({\em i.e.}\/ squarks) is, however,
proscribed by existing LHC searches in final states with jets, 
$|\ptmiss|$
\cite{Chatrchyan:2011zy,Aad:2011ib}, and possibly leptons, which put
strong (around a TeV) bounds
on both gluino and squark masses when all squarks are assumed to have
common mass. These bounds can be avoided either by relaxing the
assumption of common squark mass, keeping only the third generation
squarks light for naturalness reasons (excluding perhaps $\tilde{b}_R$) or by invoking
violation of $R$-parity.

One may also add a colour octet in the form of a light gluino and in fact
this is desirable from the naturalness viewpoint: as several authors
have stressed recently \cite{Essig:2011qg,Kats:2011qh,Brust:2011tb,Papucci:2011wy}, light, third-generation squarks are
themselves natural only if the gluino is relatively light, since the
former receive large corrections to their masses at one-loop from the
latter via the
strong interaction. (Equivalently, there are sizable contributions to
Higgs mass parameters at two-loop level from contributions involving
the gluino.) As always, it is hard to argue in absolute terms that
naturalness requires the gluino to be
within reach of the LHC\@. Nevertheless, it has been argued that
the gluino should not be more than a couple of
times heavier than the lightest third-generation squark \cite{Brust:2011tb}. 

If the gluino is within reach, then a significant boost to the squark production
cross-section can be obtained, via pair production of gluinos followed
by decay to top or bottom quarks and squarks. The enhancement is such
that there is already sensitivity at the LHC, with an ATLAS search \cite{ATLAS3}
reporting preliminary bounds on the gluino mass of up to around: 900 GeV,
for
$\tilde{g} \rightarrow \tilde{b} b \rightarrow 2b + \tilde{\chi}^0$;
650 GeV, for $\tilde{g} \rightarrow \tilde{t} t \rightarrow tb +
\tilde{\chi}^\pm$; and 750 GeV for $\tilde{g} \rightarrow 2 t +
\tilde{\chi}^0$.

These searches for gluino production followed by decays
to third-generation squarks have become a priority for the
experimental collaborations. However, the
searches focus on $R$-parity conserving scenarios, relying on the
presence of significant $|\ptmiss|$ in the final state. It is
clear from the arguments above, however, that the naturalness-based theoretical ideas
which motivate these searches do not motivate the additional assumption
of $R$-parity conservation. Thus, searches that probe scenarios that
feature $R$-parity violation, but are otherwise similar in terms of
their spectra, would seem
to be just as much of a priority, if they can be carried out.

Some of the final states that would arise in such a scenario seem to
be hopeless. Imagine, for example, pair production of gluinos
followed by decays to top squarks and quarks, followed by hadronic
decays of the top quarks and hadronic decays of the stop quarks via the $R$-parity
violating operator $W \supset U_3D_jD_k$. Such ten jet final states will
{\em a priori}\/ be
very difficult to disentangle from a background that we are currently
unable to compute accurately.\footnote{Refs.~\cite{Aad:2011qa,Chatrchyan:2011cj,Hook:2012fd} suggest
 a variety of interesting ways in which this hurdle may be overcome.}
But other final states seem much more promising. Indeed,
the presence of pair-produced gluinos not only boosts the cross
section, but also, due to the Majorana nature of the gluino, implies
an equal proportion of particle and antiparticle states, on average, in each gluino
decay. The gluinos decay to third generation quarks and
squarks, whose decays often involve leptons, leading to the
possibility of final states involving same-sign di-leptons, for which the
SM backgrounds are manageable. In the next Section, we
consider the various possibilities in more detail, showing that most
can be covered by a single, same-sign-di-lepton search.
\section{Same-sign dilepton searches \label{sec:rpv}}
In considering the final states of interest, it is useful to
divide the possibilities according to two criteria: 
(i) whether the gluino decays predominantly to stops or sbottoms (and
indeed whether these are predominantly left- or right-handed), and
(ii) whether the subsequent decay of the stop or sbottom proceeds via
the $U_iD_jD_k$ operator or the $Q_iL_jD_k$ superpotential
operator.\footnote{It is also 
  possible, in other scenarios, that cascade decay chains will
  terminate via the $L_iL_jE_k$ or $L_i H_2$ superpotential
  operators. Evidently these {\em also}\/
  result in large rates for same-sign leptons in the final states.}
Furthermore, though 
the 
right-handed bottom squark 
need not be light for naturalness, it may plausibly be present to make
up a full third generation of light squarks and so we include it in
our general discussion, distinguishing between cases in which it is, or
is not, present.
\subsection{Tops and stops}
Let us begin by analysing the case in which gluino decays produce top
quarks. If they do, then we expect same-sign
tops (or anti-tops) in half of the $\tilde{g}\tilde{g}$ events and even in the worst case
scenario, we expect same sign $e$ or
$\mu$ from top quark decays in roughly 3.3 $\%$ of events and to same sign $e$, $\mu$, or
$\tau_h$ in roughly 5.5 $\%$ events. Though these fractions are small, one must bear in mind
that the cross section for gluino production is relatively large
compared to the SM and detector backgrounds in this channel.

This worst case scenario corresponds to the lightest squark being
predominantly $\tilde{t}_R$, decaying via $U_3D_iD_j$ operators to two
jets, one of which may be a $b$ jet, illustrated in
Fig.~\ref{fig:decays-br}a. Thus, there are no additional,
isolated leptons arising from the squark decay.

In other cases, we expect same-sign leptons in a larger fraction of
events. For example if the gluino decays to a $\tilde{t}_L$ that eventually
decays via the $U_iD_jD_k$ operator, 
then the $\tilde{t}_L$ may decay in one of three ways, depending on
the Yukawa couplings and on which of the other third-generation squarks is lightest. Firstly, it may
decay to a (virtual, since it is
heavier) $\tilde{t}_R$ and a neutral Higgs boson
 as in
Fig.~\ref{fig:decays-br}b, in which case the extra number of leptons
(coming from $h^0$ decay) is expected to be negligible. Secondly, it
may decay to a $\tilde{b}_L$ and a $W$ boson
(followed by decay of the $\tilde{b}_L$ as described in the next
subsection), in which case extra leptons come from the $W$ decay.
Thirdly, if the $\tilde{b}_R$ is not too heavy, the $\tilde{t}_L$  may
decay via an off-shell
$\tilde{b}_R$ and a charged Higgs boson, as in
Fig.~\ref{fig:decays+br}a. Then there will be
an additional source of leptons coming from the decay of the
$\tilde{b}_R$,
since flavour constraints on $\lambda^\prime\prime_{lmn}$ suggest that the
up-type quark produced 
will be a top quark \cite{Allanach:1999ic,Giudice:2011ak}. Moreover, the charged
Higgs is also likely to generate a source of leptons, as we shall
shortly discuss.

Finally, the proportion of same sign leptons will be rather higher
again in all cases where the superpartner decay chains terminate via the $W
\supset Q_iL_3D_j$ operator. (Again, on the basis 
of flavour constraints upon the couplings, we expect that only the
$\lambda^\prime_{ijk}$ coupling involving
the $\tau$, {\em i.e.}\/ $j=3$ is likely to be sizable, as discussed in
\cite{Gripaios:2009dq}.) 
In such
cases, same-sign stops lead to same-sign ($\tau$) leptons, and even
opposite-sign stops will lead to same-sign di-leptons if at least one top
decays leptonically. Approximating $t$ decays such that they decay to each
flavour of lepton with a probability of $\frac{1}{9}$,
we obtain same sign leptons in roughly $\frac{7}{9}$
of gluino pair production events.
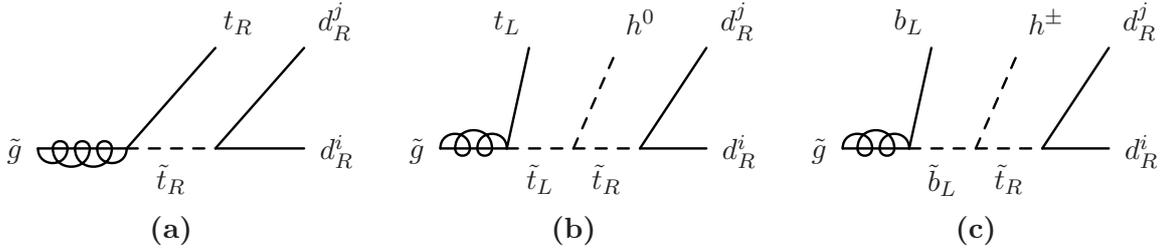
\begin{figure}[ht]
\setlength{\unitlength}{1mm}
\begin{minipage}[b]{0.3\linewidth}
\centering
\vspace{0.3cm}
\begin{fmffile}{tr} 
  \begin{fmfgraph*}(35,20)
\fmfstraight
\fmfleft{i0,i1, i2,i3}
\fmfright{o0,o1, o2,o3}
\fmftop{t1,t0,t2,t3}
   \fmflabel{$\tilde{g}$}{i1}
   \fmflabel{$d^i_R$}{o1}
 \fmf{gluino}{i1,v1}
    \fmf{dashes,label=$\tilde{t}_R$}{v1,v2}
    \fmf{plain}{v2,o1}
\fmffreeze
    \fmflabel{$d^j_R$}{o3}
    \fmflabel{$t_R$}{t2}
\fmf{plain}{v1,t2}
    \fmf{plain}{o3,v2}
 \end{fmfgraph*}
 \end{fmffile} 
\\
        {\bf (a)}
\end{minipage}
\hspace{0.5cm}
\begin{minipage}[b]{0.3\linewidth}
\centering
\begin{fmffile}{tl} 
  \begin{fmfgraph*}(35,20)
\fmfstraight
\fmfleft{i0,i1, i2,i3}
\fmfright{o0,o1, o2,o3}
\fmftop{t1,t2,t3,t4}
   \fmflabel{$\tilde{g}$}{i1}
   \fmflabel{$d^i_R$}{o1}
    \fmflabel{$d^j_R$}{o3}
    \fmflabel{$t_L$}{t2}
\fmflabel{$h^0$}{t3}
    \fmf{gluino}{v1,i1}
\fmf{dashes,label=$\tilde{t}_L$}{v1,v2}
\fmf{dashes,label=$\tilde{t}_R$}{v2,v3}
  \fmf{plain}{o1,v3}
\fmffreeze
    \fmf{plain}{v1,t2}
    \fmf{dashes}{v2,t3}
\fmf{plain}{o3,v3}
 \end{fmfgraph*}
 \end{fmffile} 
\\
        {\bf (b)}
\end{minipage}
\hspace{0.5cm}
\begin{minipage}[b]{0.3\linewidth}
\centering
\begin{fmffile}{bl} 
  \begin{fmfgraph*}(35,20)
\fmfstraight
\fmfleft{i0,i1, i2,i3}
\fmfright{o0,o1, o2,o3}
\fmftop{t1,t2,t3,t4}
   \fmflabel{$\tilde{g}$}{i1}
   \fmflabel{$d^i_R$}{o1}
    \fmflabel{$d^j_R$}{o3}
    \fmflabel{$b_L$}{t2}
\fmflabel{$h^\pm$}{t3}
    \fmf{gluino}{v1,i1}
\fmf{dashes,label=$\tilde{b}_L$}{v1,v2}
\fmf{dashes,label=$\tilde{t}_R$}{v2,v3}
  \fmf{plain}{o1,v3}
\fmffreeze
    \fmf{plain}{v1,t2}
    \fmf{dashes}{v2,t3}
\fmf{plain}{o3,v3}
 \end{fmfgraph*}
 \end{fmffile} 
\\
        {\bf (c)}
\end{minipage}
\caption{Gluino decays without
  right-handed bottom squarks in the presence of $W \supset U_3D_iD_j$, via
  (a) right-handed 
  top, (b) left-handed top, and 
  (c) left-handed bottom. Same sign leptons are
   obtained in (c) only if the charged Higgs subsequently decays to $t\overline{b}$ or
   to leptons. \label{fig:decays-br}}
\end{figure}
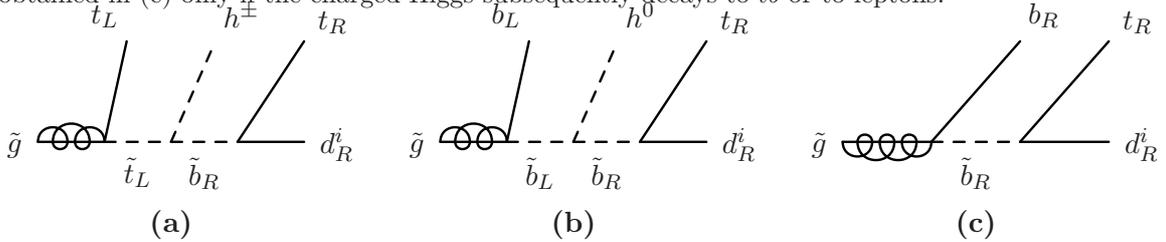
\begin{figure}[ht]
\setlength{\unitlength}{1mm}
\begin{minipage}[b]{0.3\linewidth}
\centering
\begin{fmffile}{tl2} 
  \begin{fmfgraph*}(35,20)
\fmfstraight
\fmfleft{i0,i1, i2,i3}
\fmfright{o0,o1, o2,o3}
\fmftop{t1,t2,t3,t4}
   \fmflabel{$\tilde{g}$}{i1}
   \fmflabel{$d^i_R$}{o1}
    \fmflabel{$t_R$}{o3}
    \fmflabel{$t_L$}{t2}
\fmflabel{$h^\pm$}{t3}
    \fmf{gluino}{v1,i1}
\fmf{dashes,label=$\tilde{t}_L$}{v1,v2}
\fmf{dashes,label=$\tilde{b}_R$}{v2,v3}
  \fmf{plain}{o1,v3}
\fmffreeze
    \fmf{plain}{v1,t2}
    \fmf{dashes}{v2,t3}
\fmf{plain}{o3,v3}
 \end{fmfgraph*}
 \end{fmffile} 
\\
        {\bf (a)}
\end{minipage}
\hspace{0.5cm}
\begin{minipage}[b]{0.3\linewidth}
\centering
\begin{fmffile}{bl2} 
  \begin{fmfgraph*}(35,20)
\fmfstraight
\fmfleft{i0,i1, i2,i3}
\fmfright{o0,o1, o2,o3}
\fmftop{t1,t2,t3,t4}
   \fmflabel{$\tilde{g}$}{i1}
   \fmflabel{$d^i_R$}{o1}
    \fmflabel{$t_R$}{o3}
    \fmflabel{$b_L$}{t2}
\fmflabel{$h^0$}{t3}
    \fmf{gluino}{v1,i1}
\fmf{dashes,label=$\tilde{b}_L$}{v1,v2}
\fmf{dashes,label=$\tilde{b}_R$}{v2,v3}
  \fmf{plain}{o1,v3}
\fmffreeze
    \fmf{plain}{v1,t2}
    \fmf{dashes}{v2,t3}
\fmf{plain}{o3,v3}
 \end{fmfgraph*}
 \end{fmffile} 
\\
        {\bf (b)}
\end{minipage}
\hspace{0.5cm}
\begin{minipage}[b]{0.3\linewidth}
\centering
\begin{fmffile}{br} 
  \begin{fmfgraph*}(35,20)
\fmfstraight
\fmfleft{i0,i1, i2,i3}
\fmfright{o0,o1, o2,o3}
\fmftop{t1,t0,t2,t3}
   \fmflabel{$\tilde{g}$}{i1}
   \fmflabel{$d^i_R$}{o1}
 \fmf{gluino}{i1,v1}
    \fmf{dashes,label=$\tilde{b}_R$}{v1,v2}
    \fmf{plain}{v2,o1}
\fmffreeze
    \fmflabel{$t_R$}{o3}
    \fmflabel{$b_R$}{t2}
\fmf{plain}{v1,t2}
    \fmf{plain}{o3,v2}
 \end{fmfgraph*}
 \end{fmffile} 
\\
        {\bf (c)}
\end{minipage}
\caption{Gluino decays with
  right-handed bottom squarks in the presence of $W \supset U_3D_iD_3$, via
  (a) left-handed 
 top squark, (b) left-handed bottom squark, and 
  (c) right-handed bottom squark. \label{fig:decays+br}}
\end{figure}
\subsection{Bottoms and sbottoms}
If gluinos instead decay predominantly to bottom quarks, similar
arguments apply.
 On the one hand, if the eventual $R$-parity-violating decay arises via the $Q_iL_jD_k$
operator, then we expect same-sign di-leptons (most likely $\tau$) in
at least half of events.

On the other hand, if the dominant $R-$parity violating operator is $W
\supset U_3D_iD_3$, then a $\tilde{b}_L$ produced in a gluino decay must first
decay to either a 
charged Higgs or a $W$-boson and a
(virtual) $\tilde{t}_R$ or $\tilde{t}_L$, respectively. The relative
branching fractions for these decays depend on the unknown masses of the
$\tilde{t}_R$ or $\tilde{t}_L$ , as well as on the top Yukawa
coupling. The decay involving a $W$ can obviously lead to leptons, but
this is not obviously the case for a decay involving the charged
Higgs, illustrated in Fig.~\ref{fig:decays-br}c. This, in a sense, represents
a potentially dangerous scenario, since the
$\tilde{t}_R$
will decay to two anti down-type quarks, meaning that the only possible
source of isolated leptons comes from charged Higgs decays. Thus, in
this one case out of all those we consider, a significant source of
same-sign leptons is not automatic, depending
on the details of the charged Higgs decays. However, for $m_{H^\pm} > m_t +
m_b$, decays to $t\overline{b}$ will dominate, while for lower charged Higgs
masses, decays to $\tau \nu_\tau$
will dominate, unless $\tan \beta$ is so small that the CKM
suppressed decay to $c\overline{b}$ becomes competitive. This requires
that $\tan \beta \ll \frac{m_t}{m_\tau} V_{cb} \approx 3$, which seems
highly unlikely, at least in the context of the MSSM, because of higgs
constraints from LEP.

If we follow the non-minimal possibility of a light $\tilde{b}_R$,
then the $\tilde{b}_R$ may be dominantly produced in the gluino
decay (as in Fig.~\ref{fig:decays+br}c), in which case flavour
arguments \cite{Giudice:2011ak}  suggest that a top quark (and hence leptons)
will be 
produced in the $R$-parity-violating decay. Alternatively, the gluino
may first decay to a lighter $\tilde{b}_L$, which may then decay to a
top quark via a virtual $\tilde{b}_R$ and a neutral Higgs state, as in Fig.~\ref{fig:decays+br}b.

Thus we see that in this broad range of scenarios with gluino pair
production followed by decays to top and bottom quarks and squarks (on-
or off-shell), with one $R$-parity violating coupling dominating,
same-sign di-lepton signatures are guaranteed in all but one case, {\em viz.}\/
when gluinos decay predominantly to $\tilde{b}_L$, which in turn decay
via $\tilde{t}_R$ and the $U_3D_iD_j$
operator. Even then, the resulting charged Higgs decay is likely to
proceed via top quarks or a $\tau$ lepton. 

Before going further, we pause to
discuss complications that may arise when even more
states are present in the effective theory at the weak scale. Consider what
would happen, for 
example, if Higgsinos were present in the scenario where gluinos decay
predominantly to top quarks and squarks, with the latter able to decay
via the $\lambda^{\prime \prime}_{3jk}$ coupling. Depending on the size of
this coupling and the precise spectrum, it might prove preferential
for the top squark to first decay via the supersymmetric Yukawa coupling (to
a top quark and a neutralino or a bottom quark and a
chargino). But
the resulting charginos or neutralinos cannot decay via the
$\lambda^{\prime \prime}_{3jk}$ coupling and would be forced to decay back
through to the Yukawa coupling to a virtual top squark, which in turn
would decay via the $R$-parity-violating coupling. For those decay
chains involving neutralinos an extra $t\overline{t}$ pair would be present in
each gluino decay, yet further increasing the probability of a
same-sign di-lepton event.\footnote{This increase must be weighed
  against the fact that the branching ratio for the chargino chain
  exceeds that for the neutralino chain, assuming approximate
  chargino/neutralino degeneracy.} In the most favourable case of an
event with six top quarks in the final state, the probability of a
same-sign $ee$, $\mu\mu$ or $e\mu$ rises to roughly one in four.
\subsection{Bottom tagging}
The same-sign di-lepton final states we have been discussing involve
at least two $b$ quarks and as many as six, so further leverage of the
signal compared to the background may be obtained by requesting some
number of $b$ tags. (Roughly, for a signal containing $n$
$b$ jets, and with $b$-tagging efficiency of order fifty {\em per cent},
requesting $\frac{n}{2}$ $b$ tags keeps the signal efficiency of order
a half). The most effective course of action may be to
search for different numbers of $b$ tags: up to, say, three.
\subsection{Same sign di-taus\label{sec:ditau}}
The procedure of measuring the charges of electrons and muons and the
associated uncertainties are relatively well understood. But many of
the scenarios mentioned above involve production of $\tau$s. Indeed, in cases
where the $R$-parity violating decay is via the $Q_iL_jD_k$ 
operator, we have argued, on the basis of constraints from flavour physics,
that the lepton involved is likely to be a 
$\tau$ (i.e.\ the dominant coupling has $j=3$). The latter decays to $e$ or
$\mu$ roughly one-third of the 
time, meaning that eight out of nine signal events will be lost if
we only search for same-sign pairs of electrons and muons. Similarly,
when decays occur via the $U_3D_iD_j$ operator, a lepton that comes from a top
quark decay will be a $\tau$  
roughly a third of the time, so same-sign electrons or muons arise
in just over half of di-leptonic decays.
Thus, an important question is whether we are
able to reliably measure the signs of the charges of hadronically
decaying $\tau$s in LHC events. An
answer in the affirmative is given by a recent CMS analysis
\cite{CMS} doing just that. The CMS selection employs an
algorithm (described in more detail in \cite{CMS}) in which the individual hadronic
decay modes are explicitly identified, with an efficiency that asymptotes
to 34\% for momenta above 80 GeV.

However, it is important to note that the CMS analysis requires
significant amounts of $|\ptmiss|$ in events with one or more
hadronically-decaying $\tau$s, both for triggering and for the
baseline event selection ({\em viz.} $|\ptmiss|\gtrsim 35$ GeV and
$>80$ GeV, respectively). But in scenarios in which the $R$-parity
violating decay proceeds via the $Q_iL_3D_j$ operator, there may not be
sufficient, isolated $|\ptmiss|$ in events to pass these thresholds:
whilst $|\ptmiss|$ is always present in
$\tau$ decays due to the presence of neutrinos, it may either be too
small or too closely aligned to the hadronic jet coming from a $\tau$
decay.

As a result, developing a search strategy for these scenarios is likely to
require a careful study of triggering issues and perhaps a dedicated
trigger. One possibility would be to focus on events where only one
 $\tau$ decays hadronically, using the $e$ or $\mu$ from the
other $\tau$ to trigger, possibly in tandem with a requirement on $H_T$ or other
jet activity.
\section{Current searches \label{sec:current}}
We use three current LHC searches to constrain the gluino and stop masses,
under the 
assumption that the stop or gluino decays dominantly via the $\rpv$
couplings $\lambda''_{3ij}$. 
The searches all involve same-sign di-leptons, which 
provides a good opportunity to find natural SUSY,\footnote{Supersymmetric
  like-sign di-lepton signatures have received much attention 
  in the literature for the more usual large $\ptmiss$
  searches~\cite{Baer:1991xs,Barnett:1993ea}.  
Ref.~\cite{Csaki:2011ge} also mentions the possibility of using them in an
$R-$parity violating context.} since 
backgrounds are low. 
The SUSY signal is several times larger in $W+$jets type
final states, where the $W$ decays
leptonically, and one could potentially use recent
measurements~\cite{CMS:2011aa,Aad:2012en} of the 
properties of such final states to bound the new physics models. However, the
SM
backgrounds are expected to be much larger such that the model coverage 
would be smaller than for same-sign di-lepton signatures. Moreover,
$W+$jets searches may not provide coverage of scenarios where the
leptons arise from the $Q_iL_jD_k$ operator  instead of from top quark
decays.

The dominant backgrounds to like-sign di-lepton production include $t \bar t W$
production, 
and ``fake leptons'',
where  jets can yield isolated leptons from unidentified photon conversions,
muons from meson decays in flight, heavy flavour decays, or other detector
effects~\cite{CMS2, ATLAS2}. We do not simulate such backgrounds, since the
experimental publications have already taken them into account and
have provided
bounds on new physics contributions to the cross section after cuts.
Since we do not make
use of any $b$ tags, our re-interpretation of the searches does not depend
upon how many of the $\lambda''_{3ij}$ are non-zero, nor on the values of $i$ and
$j$. For the simulations, we have explicitly chosen only one non-zero
weak-scale baryon-number violating coupling, with $i=2$ and $j=3$. 

\begin{figure}\begin{center}
\unitlength=1in
\begin{picture}(4,4)(0,0)
\put(-0.85,3.8){\includegraphics[width=4in,angle=270]{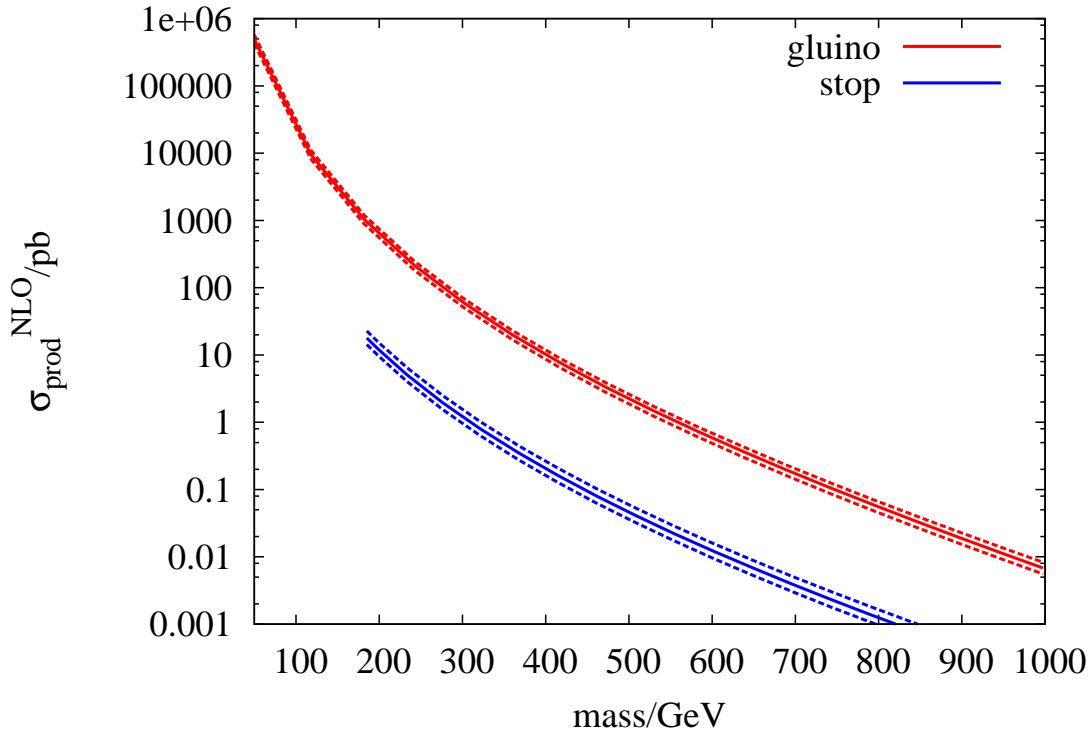}}
\end{picture}
\caption{Stop and gluino total production cross-sections at a 7 TeV LHC
  calculated at NLO by {\tt PROSPINO2.1}. The
  dashed curves show the variations due to changes in the renormalisation scale
 by a factor of two. \label{fig:sigma}}
\end{center}
\end{figure}

We shall approximate experimental searches by simulating the signal events
with {\tt HERWIG++-2.5.2}~\cite{Gigg:2007cr,Bahr:2008pv,Gigg:2008yc} and
reconstructing jets with {\tt fastjet-3.0.1}~\cite{Cacciari:2011ma}, using
cuts identical to those used in the experimental analyses. For the most constraining
search, we include detector corrected efficiencies, whereas for the others,
these effects 
are neglected, since they are expected to only weaken the exclusion.
We calculate the MSSM spectrum with {\tt
  SOFTSUSY2.4.3}~\cite{Allanach:2001kg,Allanach:2009bv}, passing the
information on to the event generator via the SUSY Les Houches
Accord~\cite{Skands:2003cj}. Sparticle decays are calculated using
{\tt
  PYTHIA6.4.25}~\cite{Sjostrand:2006za}.  
Our simplified model has all sparticles other than the right-handed stops 
and gluinos being heavy. The only SUSY
production processes at the 7 TeV LHC are then 
di-stop production and di-gluino production. We plot the total production
cross-sections of di-gluinos and di-stops in Fig.~\ref{fig:sigma}, calculated
to next-to-leading order in QCD with {\tt
  PROSPINO2.1}~\cite{Beenakker:1996ch,Beenakker:1997ut}.  We use the central
next-to-leading order
(NLO) production cross-sections for our estimates of the exclusion
power of various
searches. 
We see large
cross-sections for light gluinos, and expect that current LHC  data of
$\sim$5000 pb$^{-1}$ of 7 TeV $pp$ collisions should constrain them. 
\begin{figure}\begin{center}
\unitlength=1in
\begin{picture}(4,3)(0,0)
\put(-0.85,3.6){\includegraphics[width=4in,angle=270]{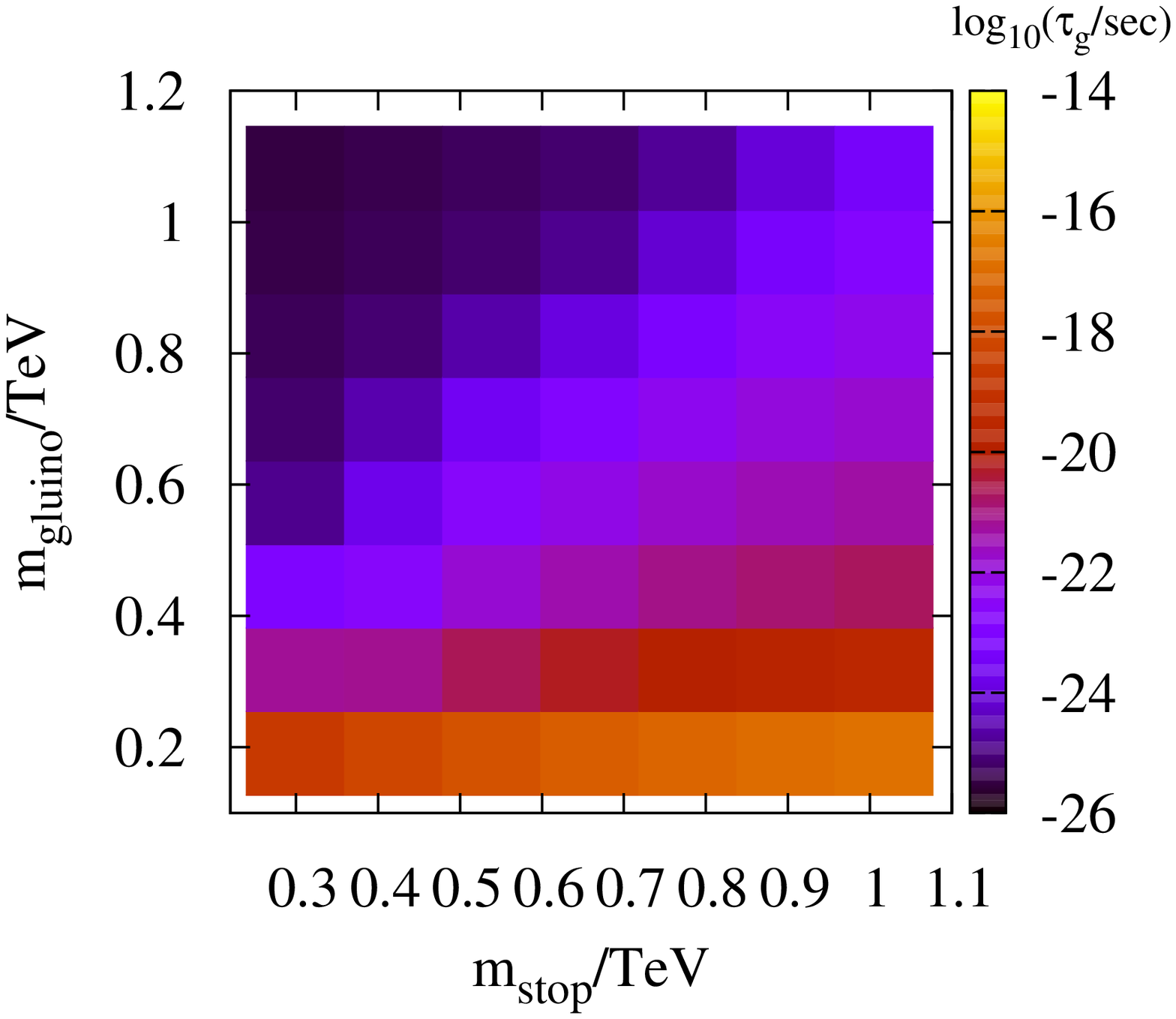}}
\put(0.1,3.05){\includegraphics[width=2.7in,angle=270]{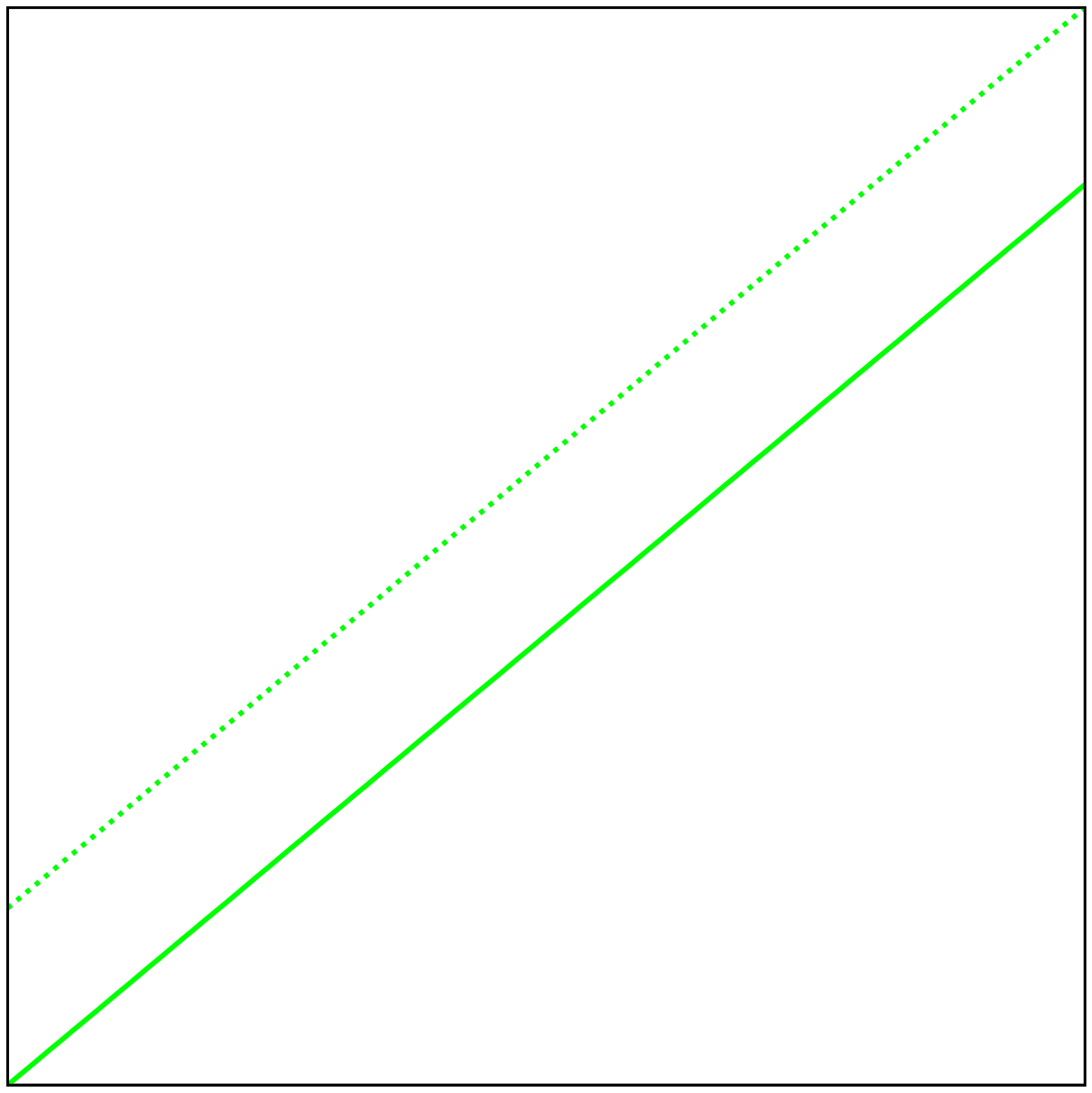}}
\end{picture}
\caption{Gluino lifetime $\tau_g$ for $\lambda^{\prime\prime}_{323}=1$. The green unbroken contour
  displays the line of equal mass, where the broken line shows $m_{\tilde
    g}=m_t + m_{\tilde t}$.  \label{fig:life}}
\end{center}
\end{figure}
We show the gluino lifetime for a particular value of $\lambda''_{323}=1$ in
Fig.~\ref{fig:life}, as calculated by {\tt
  PYTHIA6.4.25}~\cite{Sjostrand:2006za}.  
This ignores decays through an off-shell top and finite-width effects
(these should be negligible, however, except very close to the broken
line in Fig.~\ref{fig:life}. To a good approximation, the gluino
lifetime above the broken line (where 
$\tilde g   \rightarrow t^*\tilde t,\ t {\tilde t}^*$ and both the
stop and top
are on-shell) is insensitive to the value of 
$\lambda''_{3ij}$, because the decay is mediated by a SUSY gauge
interaction. The lifetime is then governed by the strong 
coupling $\alpha_s$   and by the masses of the gluino, top and stop.   
Below the broken line, $\tilde g \rightarrow
t \bar s \bar b$ through an off-shell stop (plus
conjugate decay products). In this case, $\tau_g   \propto
1/|\lambda^{\prime\prime}_{323}|^2$. The largest lifetime of $\sim 10^{-15}$
seconds corresponds to a decay length of $c \tau_{\tilde g}=3\times
10^{-7}$m, so all decays are prompt in the parameter region shown in
the Figure. If, however, $\lambda^{\prime\prime}_{323}$ were to be 
reduced by a couple of orders of magnitude, one would obtain gluinos
that would form
millimetre length (or longer) $R-$hadron tracks~\cite{Kraan:2005ji},
terminating in a decay into jets or jets and a lepton. We do not
consider this interesting possibility further here.

It is important to consider the lepton isolation criteria, since they are the
main tool used to reduce the fake lepton backgrounds. For some test lepton $l$,
each of 
the three searches we consider defines a quantity $p_T^{cone}$, which
must satisfy the following
inequality in order for a lepton in the fiducial region of the detector to be
considered isolated: 
\begin{equation}
p_T^{cone}(p_T^{min}, \Delta R) < \mbox{max} (I_{iso}
p_T(l),\ M_{iso}), \label{isoCrit} 
\end{equation}
where $p_T^{cone}$ is defined as the sum of $p_T$ of all tracks with
$p_T>p_T^{min}$ 
in a cone around the lepton axis described by $\Delta R =
\sqrt{(\eta-\eta(l))^2 + (\phi-\phi(l))^2}$, where $\eta$ is the
pseudo-rapidity and $\phi$ the azimuthal angle, measured in radians. 

\subsection{Test point \label{sec:test}}
We shall define a specific point in parameter space to illustrate various
properties of signal events in LHC collisions. Our point has
$m_{{\tilde t}_R}=500$, $M_3=478$ GeV, $\lambda''_{323}=1$ and all other
sparticles heavy. For convenience, we will set other 
$\lambda''$ couplings to zero in 
our simulations, but our results would be identical were we to include
them\footnote{We note that even if only one $\lambda''_{ijk}$ coupling were
  non-zero 
  in the interaction eigenbasis, rotations to the quark mass eigenbasis will
  induce others.}. 
This point yields 
\begin{equation}
m_{\tilde g}=588\mbox{~GeV}, \qquad
m_{{\tilde t}_1}=581 \mbox{~GeV}.
\end{equation}
Gluinos then promptly decay through off-shell stops into $tsb$ or $\bar t \bar s
\bar b$, with an equal branching ratio for each channel. 
We summarise the sparticle decays in
Table~\ref{tab:decays}.
The total cross-section for production of supersymmetric particles before cuts
is $\sigma_{NLO}=681$ fb for the default renormalisation scale. 
\begin{table}\begin{center}
\begin{tabular}{|cc|cc|}\hline
Decay & Branching ratio  & Decay & Branching ratio\\ \hline
$\tilde g \rightarrow \bar t \bar s \bar b$ & 0.5 &
$\tilde g \rightarrow tsb$ & 0.5 \\
${\tilde t}_1 \rightarrow \bar s \bar b$ & 1 &
${\tilde t}_1^* \rightarrow s b$ & 1 \\
\hline \end{tabular}
\caption{Decays of stops and gluinos for the test point. \label{tab:decays}}
\end{center}
\end{table}

\subsection{ATLAS di-muon search}
We first consider an ATLAS search \cite{Aad:2012cg}, which uses prompt
like-sign muon 
pairs. This search does not require large $|\ptmiss|$, and is quite
inclusive.  This fits our
expectations for signal events, which do not contain a stable, lightest,
supersymmetric particle to carry off $|\ptmiss|$. 
ATLAS looked at 1.61 fb$^{-1}$worth of integrated luminosity at the 7 TeV
LHC, requiring two 
isolated muons with identical charges and 
$p_T>20$ GeV.
Since the data were found to agree well with SM backgrounds, upper limits were
placed on new physics cross-sections leading to like-sign di-muons. 
For this analysis, a muon with $p_T(\mu)$ is defined to be isolated if
Eq.~\ref{isoCrit} is satisfied, 
with $p_T^{min}=1$
GeV, $\Delta R = 0.4$, $I_{iso}=0.08$ and 
$M_{iso}=5$ GeV. In our simulations, the $p_T^{min}$ cut is implemented for
any charged hadron in the final state (i.e.\ all charged hadrons are assumed
to form a track). 
The fiducial region for muons is $|\eta|<2.4$. Four different signal regions
were defined, each by a different lower cut on $m_{\mu \mu}$. 
ATLAS places upper bounds on isolated like-sign di-muon production above
backgrounds and within the cuts $\sigma_{SS \mu\mu}^{95}$
as shown in the final column of Table~\ref{tab:ssmu}. 
As can be seen from the table, the acceptance $A$ of signal events is
very low, due mainly to the small branching ratios of top pairs into
isolated di-leptons. 
We simulate SUSY signal events, calculating the proportion of gluino
pairs that yield isolated like-sign di-muon pairs past the ATLAS cuts. We find
that 
this search does not yield the most restrictive bounds upon the parameter
space, and so we do not complicate the analysis and further weaken  the bounds
by performing a detector simulation, or by correcting for muon
efficiencies.\footnote{Muon efficiencies are high: from $Z-$boson decays, they
  range from 87$\%$ to 97$\%$~\cite{Aad:2012cg}, depending upon
  $p_T$. However, the muon $p_T$ spectra depend upon the new physics model and
  the ATLAS publication does not parametrise the efficiencies as a function
  of $p_T$, thus we are not able to reliably take them into account. 
}
\begin{table}\begin{center}
\begin{tabular}{|c|cccc|}\hline
Signal Region & $m_{\mu \mu}$/GeV & $\sigma_{SS\mu\mu}^{test}$/fb & $A/10^{-3}$ & $\sigma_{SS\mu\mu}^{95}$/fb  \\ \hline 
ATLAS$\mu\mu1$ & $>$15  & 12 & 1.3  & 58\\
ATLAS$\mu\mu2$ & $>$100 & 7.5 & 0.86 & 16\\
ATLAS$\mu\mu3$ & $>$200 & 2.1 & 0.29  &  8.4\\
ATLAS$\mu\mu4$ & $>$300 & 0.41 & 0.077  & 5.3\\ \hline
\end{tabular}
\caption{The ATLAS same-sign di-muon  analysis search regions. 
 The expected signal cross-section past cuts predicted our test model
 $\sigma_{SS\mu\mu}^{test}$/fb 
 is shown, as well as the acceptance of the selection $A$. $A$ is defined
 to  be the number of simulated supersymmetric events past cuts divided by the
 total number of simulated supersymmetric events. 
 In the last column, we show
 the~\cite{Aad:2012cg} 95$\%$    
 CL$_s$ upper bound on non-SM cross section past cuts found by
 ATLAS $\sigma_{SS\mu\mu}^{95}$/fb.  
  \label{tab:ssmu}}
\end{center}
\end{table}
While simulating same-sign di-muon signals, we force the tops to decay via
$e,\mu$ or $\tau$ in order to get better Monte Carlo statistics, taking
the associated correction factor into account.  
For each point considered in parameter space, we simulate 10~000 SUSY signal events.

\subsection{ATLAS same-sign di-lepton, jets and missing transverse momentum
  search} 

\begin{table}\begin{center}
\begin{tabular}{|c|ccccc|}\hline
Signal Region & $|\ptmiss|$/GeV & $m_T(l_1)/GeV$ &  
$\sigma^{test}_{SSll}/$fb & $A/10^{-3}$   & $\sigma_{SSll}^{95}/$fb\\ \hline 
ATLAS$ll1$ & $>150$ & $>0$ &   0.67  & 1.0 & $1.6$ \\
ATLAS$ll2$ & $>150$ & $>100$ &  0.40 & 0.6  & $1.5$ \\ \hline
\end{tabular}
\caption{ATLAS same sign-di lepton analysis search regions. We show the cuts,
  and the expected cross-section past cuts
  predicted by our test point over SM backgrounds
  $\sigma_{SSll}^{test}$, as well as the acceptance of the selection $A$. $A$
  is defined 
 to  be the number of simulated supersymmetric events past cuts divided by the
 total number of simulated supersymmetric events. In the  
  last column, we show 
  the~\cite{ATLAS2} 95$\%$    
  CL$_s$ upper bound on the cross-section past cuts surplus to those from the
  SM coming from the search, $\sigma_{SSll}^{95}$. 
  \label{tab:atlasDiResults}}
\end{center}
\end{table}
In this search, the ATLAS experiment analysed 2.06 fb$^{-1}$ of integrated luminosity of LHC
collisions collected at 7 TeV centre of mass energy~\cite{ATLAS2}, looking for
events with same-sign leptons (electrons or muons) each with $p_T>20$ GeV,
large missing 
transverse momentum $|\ptmiss|>150$ GeV, and at least four jets, each with
transverse 
momenta exceeding 50 GeV. Electrons
were required to be within $|\eta|<2.47$ and the isolation criteria followed
Eq.~\ref{isoCrit} with $M_{iso}=0$, $I_{iso}=0.1$, $\Delta R=0.2$
$p_T^{min}=0$.  
For muons, $|\eta|<2.4$, $M_{iso}=1.8$ GeV, $I_{iso}=0$, $\Delta R=0.2$ and
$p_T^{min}=0$. 
Jets were defined using the anti-$k_T$
algorithm with distance parameter $R=0.4$, requiring $p_T>20$ GeV and
$|\eta|<4.5$. 
$\ptmiss$ was defined to be the vector sum of the transverse momenta of jets and
leptons, plus calorimetric energy clusters not belonging to reconstructed
objects. 
The analysis defines one signal region  which had an additional cut on the
transverse mass $M_T=\sqrt{2\ |\ptmiss|\cdot p_T(l)\cdot  [1 - \cos \Delta
    \phi(l, \ptmiss)]}$ of the hardest lepton $l$ of the same-sign pair. 
No events past cuts were observed for either signal region, allowing ATLAS to
place a 95$\%$ $CL_s$ upper limit on an extra 
component of  cross-section past cuts as shown in
Table~\ref{tab:atlasDiResults}. 
While simulating these ATLAS same-sign lepton signals, we force the tops to
decay via 
leptons (or anti-leptons) of any flavour in order to get better Monte Carlo
statistics, taking the associated correction factor into account in the
efficiencies. 
For each point considered in parameter space, we simulate 100~000 SUSY signal
events. It turns out that these searches also did not yield the most stringent
bounds upon our model. 

\subsection{CMS same-sign di-lepton, jets and missing transverse momentum
  search \label{sec:CMSsamesign}} 
The CMS experiment analysed 0.98 fb$^{-1}$ of integrated luminosity of LHC
collisions collected at 7 TeV centre of mass energy~\cite{CMS}. The leptons
were required to be within $|\eta|<2.4$ and the isolation criteria followed
Eq.~\ref{isoCrit} with $M_{iso}=0$, $I_{iso}=0.15$, $\Delta R=0.3$,
$p_T^{min}=0$. The inclusive di-leptons search  
used here 
was defined to have a baseline selection of electron $p_T>10$ GeV and muon
$p_T>5$ GeV. 
Jets were defined using the anti-$k_T$
algorithm with distance parameter $R=0.5$, requiring $p_T>40$ GeV and
$|\eta|<2.5$. Lower cuts were placed on $H_T$, defined to be the scalar sum of
jet $p_T$s that have $\Delta R>0.4$ to the closest isolated lepton passing all
other requirements. Lower cuts were also placed upon $|\ptmiss|$, which the
experiment defines using the particle flow technique. In our signal
simulation, we define $\ptmiss$ to be the vector sum of jet and isolated
lepton transverse momenta. 
CMS defined several search regions for their analysis, based on different
cuts. The search regions for the inclusive di-leptons baseline selection (for
which a  description of the detector efficiencies were given) are displayed in
Table~\ref{tab:cmsResults}. 
\begin{table}\begin{center}
\begin{tabular}{|c|ccccc|}\hline
Signal Region & $H_T$/GeV & $|\ptmiss|$/GeV & $N_{ll}$/fb &
$A \times \epsilon/10^{-3}$ & $N_{ll}^{95}$\\ \hline 
CMS$ll1$ & $>$400 & $>$120 & 2.4 & 3.5 & $<$3.7\\ 
CMS$ll2$ & $>$400 & $>50$  & 4.6 & 6.8 & $<$8.9\\
CMS$ll3$ & $>$200 & $>$120 & 2.5 & 3.7 & $<$7.3 \\ \hline 
\end{tabular}
\caption{Number of  events past cuts for the CMS same
  sign-di lepton analysis 
  $N_{ll}$ predicted by our test point over SM backgrounds, and
  acceptance $A$ times efficiency $\epsilon$ of the signal selection, for the
  test point. In the
  last column, we show 
  the~\cite{CMS} 95$\%$    
  CL$_s$ upper bound on the number of events surplus to those from the
  SM quoted by CMS, $N_{ll}^{95}$. 
The quoted acceptance $A$ is defined to be the expected number of
supersymmetric events past cuts divided by the total number of supersymmetric
events for the test point. The 
efficiency  $\epsilon$ is calculated as described in the text.
  \label{tab:cmsResults}}
\end{center}
\end{table}
CMS give approximate fitted formul\ae~for the efficiencies $\epsilon_e(p_T^e)$,
$\epsilon_\mu(p_T^\mu)$, $\epsilon_{H_T}(H_T)$, 
$\epsilon_{|\ptmiss|}(|\ptmiss|)$ of the detection of electrons, the detection
of muons, the $H_T$ cut, and the $|\ptmiss|$ cut, respectively. We take these
into account for each of our signal events by recording an efficiency for each
event 
of $\epsilon_{H_T}(H_T) \times \epsilon_{|\ptmiss|}(|\ptmiss|)$ multiplied by
the two relevant lepton efficiencies, 
 if the event yields like-sign isolated leptons within 
the fiducial region, and $H_T>200$ GeV and $|\ptmiss|>30$ GeV, {\em
  i.e.}\ where the CMS 
parametrisation applies. This procedure should take detector effects into
account at the few tens of percent level. This is sufficient for our purposes,
despite the fact that it misses possible correlations
between the different variables. 
While simulating same-sign di-muon signals, we force the tops to decay via
leptons (or anti-leptons) of any flavour in order to get better Monte Carlo
statistics, taking the associated correction factor into account. 
For each point considered in parameter space, we simulate 10~000 SUSY signal
events. 

\subsection{Model exclusion limits}

\begin{figure}[ht!]\begin{center}
\subfigure{\includegraphics[angle=270,width=\textwidth]{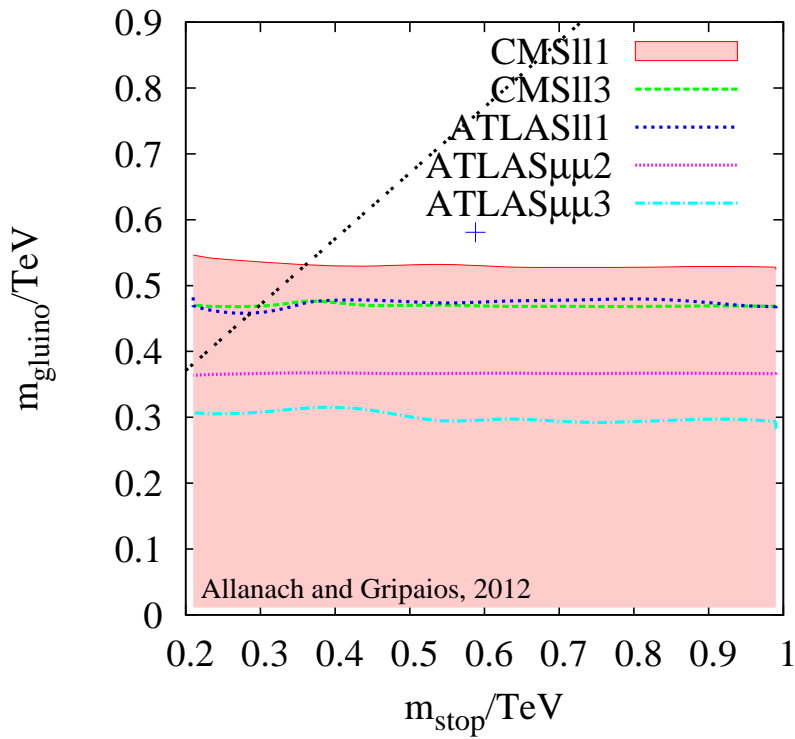}} 
\caption{95$\%$ exclusions from the CMS di-leptons $\ptmiss$ search
  (CMSll1,3),  
the ATLAS di-leptons $\ptmiss$ search (ATLASll1) and the
ATLAS same-sign di-muons search (ATLAS$\mu\mu$2,3). 
The signal regions of each
analysis, corresponding to the number are defined in
Tables~\ref{tab:ssmu},\ref{tab:atlasDiResults},\ref{tab:cmsResults}. The cross
shows the location of our test point.  We have not combined the different
exclusion regions, but have merely shaded the area excluded by the most
constraining search. Above and to the left of the black dashed diagonal line,
gluino decays are through on-shell stops, whereas underneath and to the right
of it, gluino decays are through off-shell stops. 
\label {fig:exclusion}}
\end{center}\end{figure}
We present the model exclusion limits from the ATLAS di-muon and both
di-lepton-jets-$|\ptmiss|$ analyses in Fig.~\ref{fig:exclusion}.
One of the CMS di-lepton-jets-$\ptmiss$ search regions yields the strongest
bounds upon our scenario, one of the search regions ruling out gluino 
masses less than 550 GeV. The CMS$ll2$ curve lies on top of the
CMS$ll$1 exclusion curve, and so we do not display it in the figure.
We have neglected some search regions either because they are weaker than the
search regions shown for the particular search in question.
We note that, in order to 
combine different search regions, one should apply
the {\em expected}\/ most constraining search region at each parameter point
(not the 
observed most constraining search). This procedure removes the {\em a
  posteriori}\/ statistical bias 
of just choosing the most restrictive limit in parameter space. 
However, not all of the searches supplied the expected limits from each signal
region 
and so we were unable to combined the exclusion limits in this way. Instead,
we just present the limits separately.

\begin{figure}[ht!]
\centering
\includegraphics[angle=270,width=14cm]{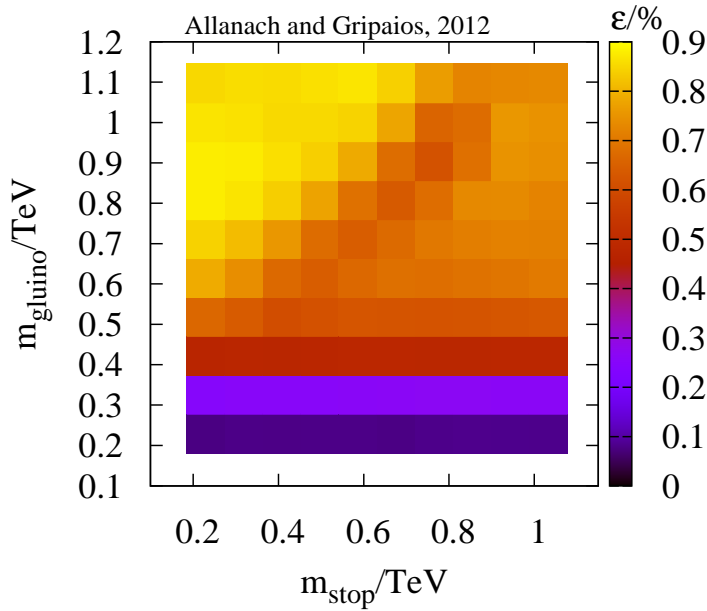}
 \caption{Signal efficiency for the the CMS di-lepton $|\ptmiss|$
   analysis. \label{fig:eff}} 
\end{figure}

In Figure~\ref{fig:exclusion}, we see that the exclusion
limits are approximately independent of the stop mass. This is expected, since
the signal production cross-section for gluino pairs is insensitive
to the stop mass. In the top-left hand portion of the plane, the gluino decays
through on-shell tops, which could have a mild effect on acceptances for the
di-lepton searches, where jet cuts are applied. 
On the other hand, the CMS$ll1$ exclusion does show a mild, but non-trivial
dependence upon  the stop 
mass. To the extreme left of the curve, gluinos may decay through on-shell
stops, but for $m_{\tilde t}> 400$ GeV, the gluino decays are three body. 
Three-body decays share the gluino's mass energy between the $tsb$ decay
products, leading to differences in the jet kinematics and $\ptmiss$.
We see this effect upon the signal efficiency in Fig.~\ref{fig:eff}. The
signal efficiency is defined as the fraction of simulated SUSY events which
yield like-sign isolated di-leptons in the fiducial region (where the leptons
satisfy the basic 
minimum $p_T$ requirements) and which also satisfy $H_T>200$ GeV,
$|\ptmiss|>30$ GeV, multiplied by
the product of detection efficiencies defined
in \S~\ref{sec:CMSsamesign}. This definition includes the leptonic
branching ratios of top pairs within the efficiency on the plot.  
The efficiency is 
rather low, less than a percent irrespective of gluino or stop mass. We
suggest below how the efficiency may be improved. 
For the sake of brevity and because detector effects have not been included, we do not show the cut efficiencies for the other, less
constraining searches.
One sees that at low values of $m_{\tilde g}$, the efficiencies are low
because of lower $p_T$ jets (however here, the production cross-sections are
extremely 
large, as Fig.~\ref{fig:sigma} shows). 

\section{Suggestions for future searches \label{sec:future}}
\begin{figure}[!ht]\begin{center}
\subfigure[$N_{SS l}$]{\includegraphics[width=0.4\textwidth]{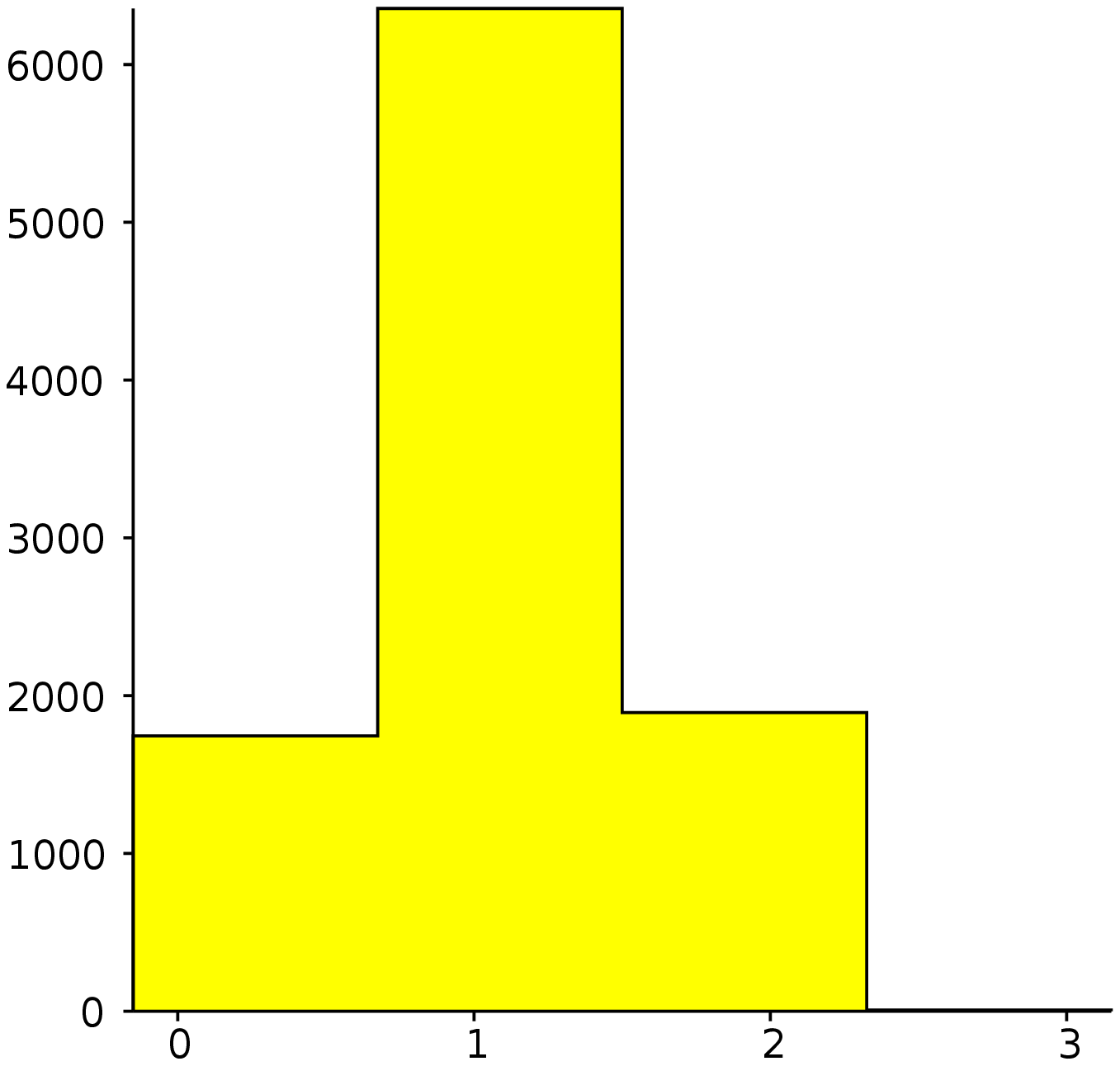}}
\hspace{2ex}
\subfigure[$p_T(l_1)/GeV$]{\includegraphics[width=0.4\textwidth]{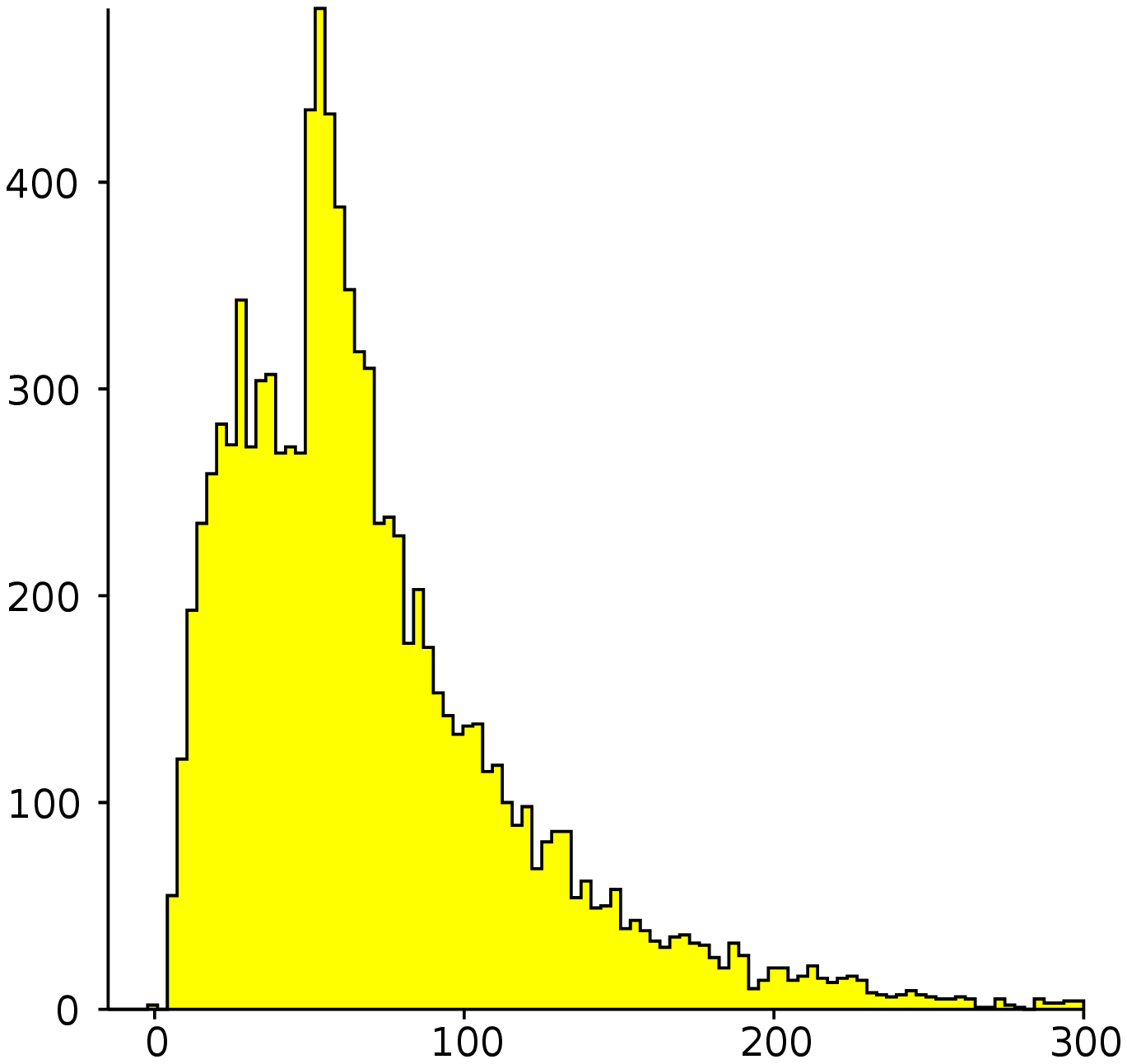}}

\subfigure[$p_T(j_1)/GeV$]{\includegraphics[width=0.4\textwidth]{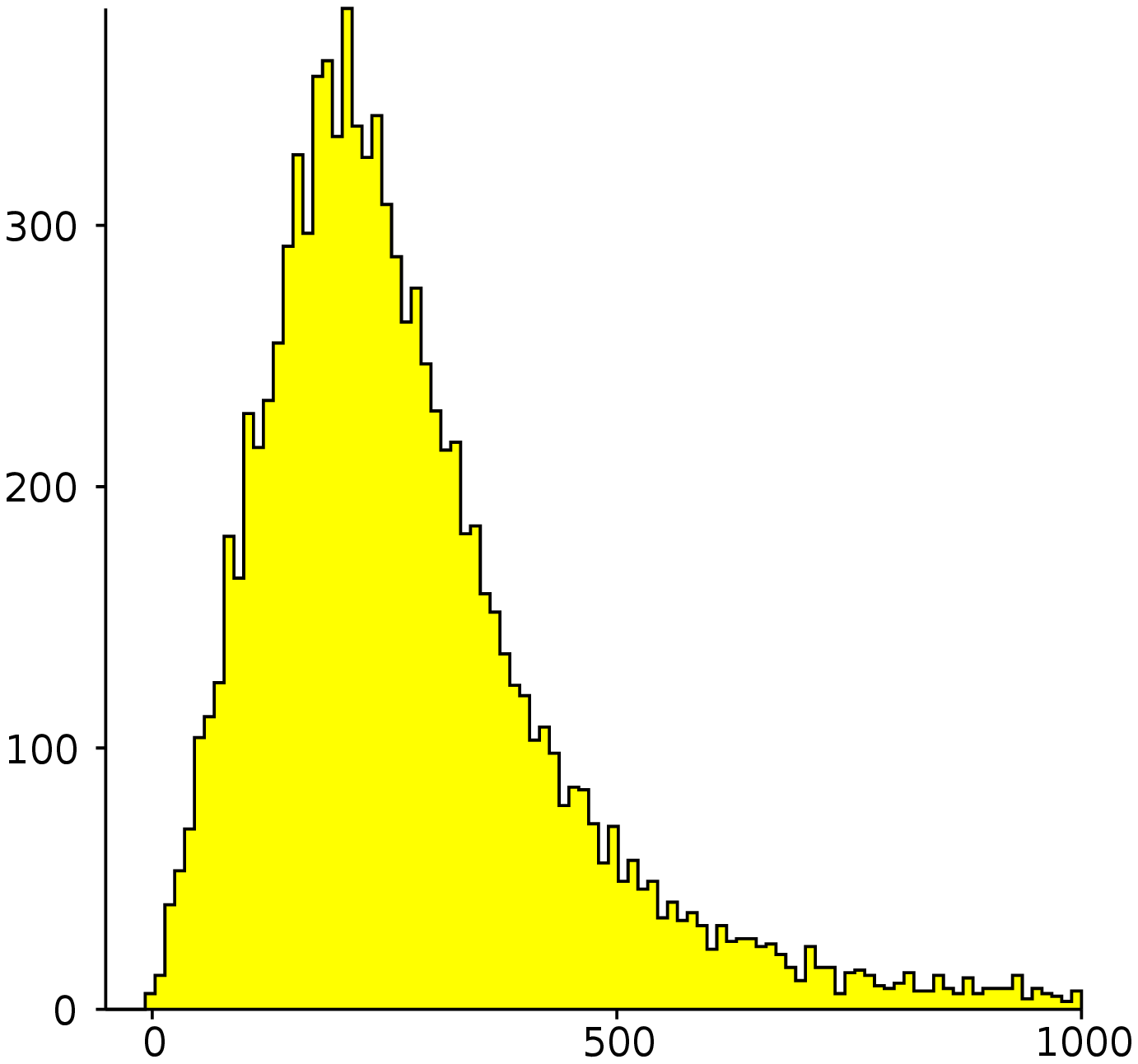}}
\hspace{2ex}
\subfigure[$N_J(p_T>20)$ GeV]{\includegraphics[width=0.4\textwidth]{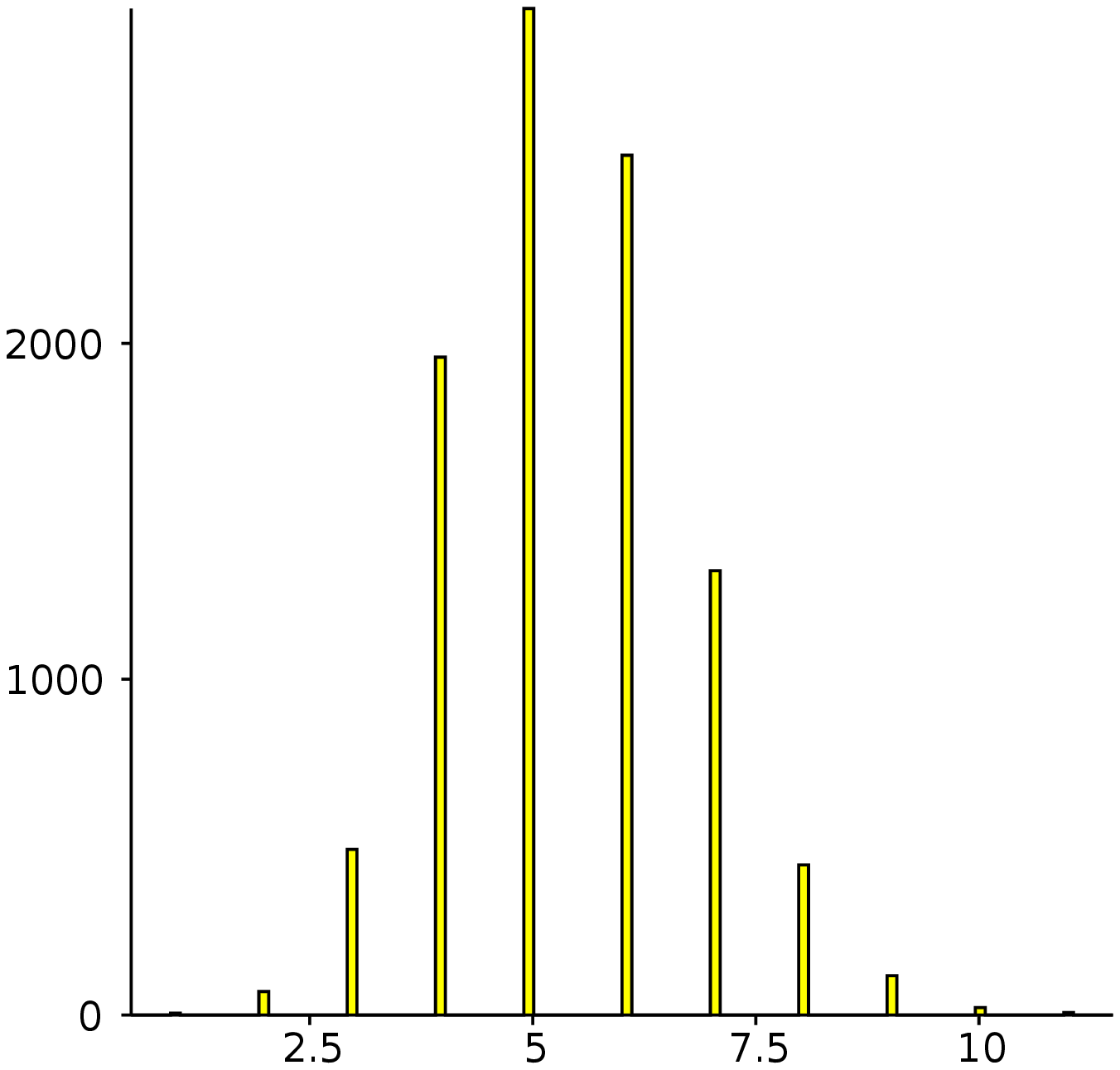}}

\subfigure[$|\ptmiss|/GeV$]{\includegraphics[width=0.4\textwidth]{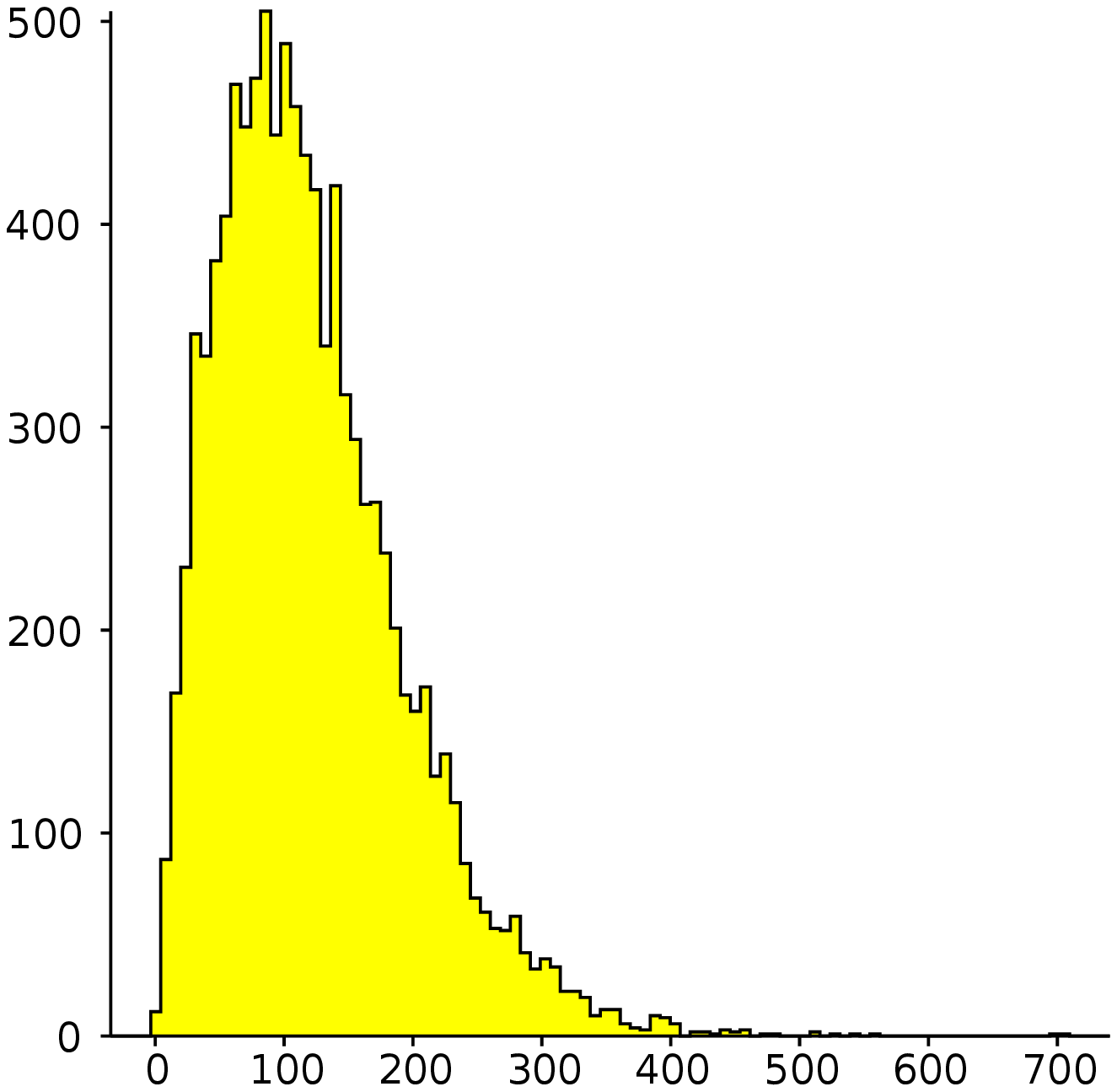}}
\hspace{2ex}
\subfigure[$H_T/GeV$]{\includegraphics[width=0.4\textwidth]{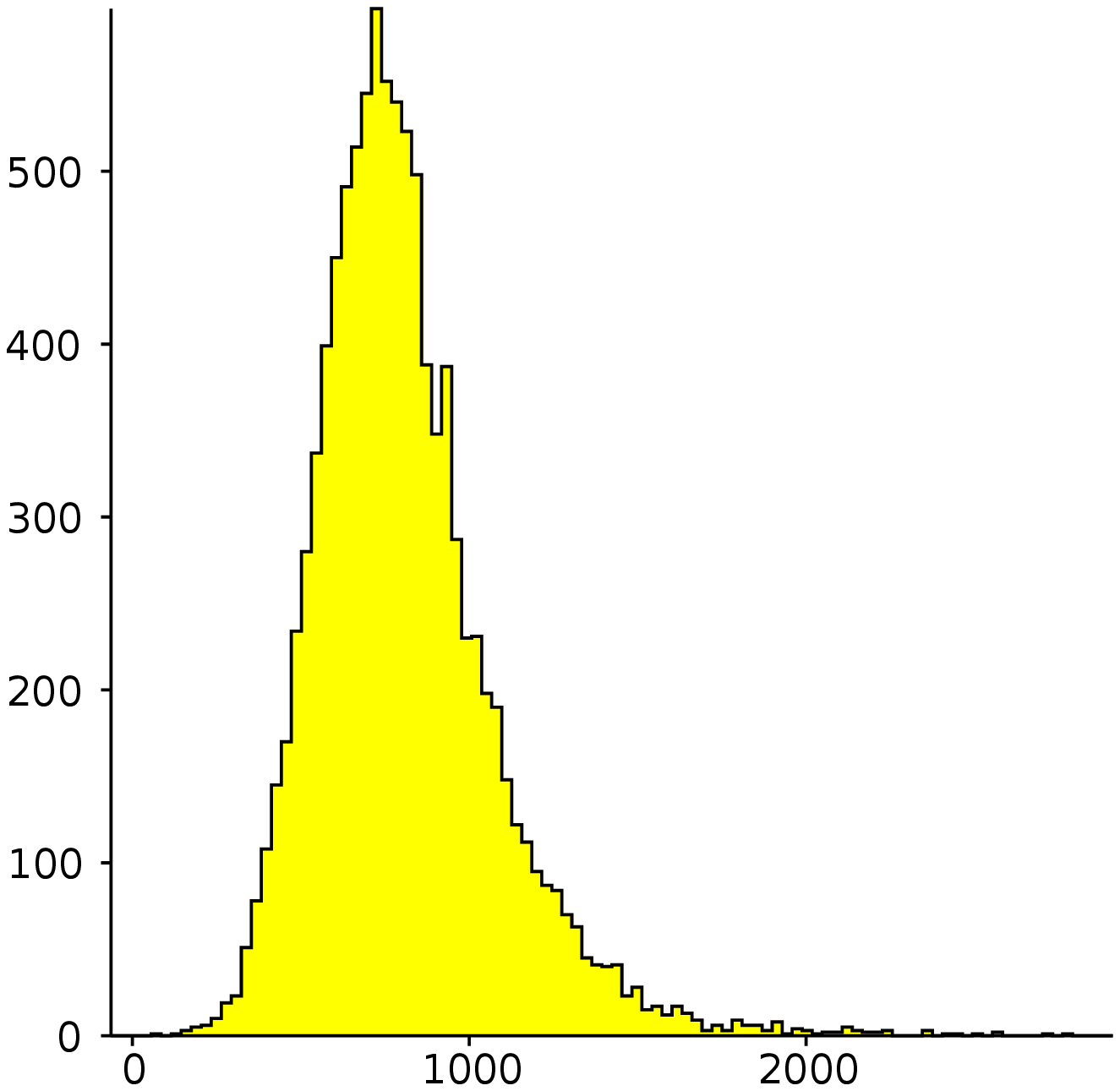}}

\caption{Distributions of quantities of interest for our test
  point $m_{\tilde g}=588\mbox{~GeV}$, $m_{{\tilde t}_1}=581 \mbox{~GeV}$, for
  supersymmetric signal events. In (a), we show the number of
  {\em isolated}\/ leptons passing minimum $p_T$ cuts (where, if there are two,
  they must have the same sign). We use the CMS
    analysis' lepton isolation criteria.
  In (b), we show the hardest isolated lepton's $p_T$. Otherwise, the only
  cuts are 
  that jets have $p_T>20$ GeV, and jets and leptons lie within the fiducial
  region of the detector. Other detector effects are not taken
  into account.   
  \label{fig:kin}
}
\end{center}\end{figure}

The searches we have considered were designed with rather different
models of new physics in mind and it is of interest to consider ways
in which they might be optimised for the models discussed here.
While a definitive answer will also require dedicated background and full
detector simulations, one can still obtain some ideas for directions for 
future study by examining properties of the signal. 
Thus, we begin by showing in Fig.~\ref{fig:kin} various distributions taken
from our 
simulated 7 TeV LHC SUSY events for the test point described in
\S\ref{sec:test}. Each quantity is defined to be
at the generator level, {\em i.e.}\ detector effects are not taken into
account.
$p_T> 20$ GeV cuts on the anti-$k_T$ ($R=0.5$) jets  are applied, as well as 
$p_T(e)>10$ GeV, $p_T(\mu)>5$ GeV, 
and fiducial volume cuts on jets and
leptons, as in \S\ref{sec:CMSsamesign}.

We see from Fig.~\ref{fig:kin}a that the
majority of events do not have two isolated like-sign di-leptons passing the
$p_T$ cuts: one lepton is often lost due to it not being isolated, or being
outside the fiducial volume of the detector. The
efficiency to select same-sign di-leptons could certainly be
improved by lowering the lepton $p_T$ cut and it is also possible that an
improvement might be obtained by modifying the lepton isolation criterion. 

Several of the existing searches also require a significant amount of missing
energy in events. For the particular model and test point considered here, a
significant amount of $|\ptmiss|$ is present (see Fig.~\ref{fig:kin}e),
though a cut at 120 GeV nevertheless removes roughly half of the
events. Unfortunately, the possible improvement in sensitivity that might be
gained by lowering (or indeed increasing) the $|\ptmiss|$ cut is difficult
to gauge from the existing searches. The CMS search, for example,
contains
search regions which are identical except for the $|\ptmiss|$ cuts applied,
but one search region is subject to an over-fluctuation in the data whilst the
other is subject to an under-fluctuation; without knowing the limit that CMS
expected to set on a signal cross-section in each region, one cannot pinpoint
the effect of varying the $|\ptmiss|$ cut. 

In any case, it is clear that there are scenarios in which a large amount of $|\ptmiss|$ cannot be expected. In particular, in cases where superpartners
eventually decay via the $Q_iL_jD_k$ operator, $|\ptmiss|$ comes only
from neutrinos from  $\tau$ decays and these may not generate
significant $|\ptmiss|$ (or indeed $|\ptmiss|$ that is
sufficiently isolated from hadronic activity to be considered
`clean'). 

Lowering the $|\ptmiss|$ and lepton $p_T$ cuts does, of course, have the
negative impact of increasing the background contribution. This could be
mitigated by increasing the cuts on other quantities. For our test point, we
see that the hadronic activity, as measured by the $p_T$ of the hardest jet
(Fig.~\ref{fig:kin}c), by the number of jets with $p_T>20$ GeV
(Fig.~\ref{fig:kin}d), or by $H_T$ (Fig.~\ref{fig:kin}f), is typically large
in signal events. Then again, one can imagine scenarios in which the jets,
though numerous, are rather soft. For example, this will be the case if the
gluino decay proceeds 
via a long chain of (possibly virtual) states, such as when squarks 
first decay to charginos or neutralinos, which in turn decay via
virtual squarks and an $R$-parity violating operator $U_iD_jD_k$. 

Similar arguments apply to $b$ tags. We have argued that one can
expect between two and six $b$ jets in signal events; the best
strategy to exploit this could be to search in regions with
differing numbers of required $b$ tags, so as to give maximum coverage in model
space. 

Most of these effects should be relatively straightforward to
analyse and implement. More challenging is the issue of how to optimise one's
searches for same-sign dilepton final states involving one or more
hadronically-decaying $\tau$. We have argued that these may be
dominant in some models. The approach described in 
\cite{CMS} acts as a proof-of-principle that searches using hadronic $\tau$
decays can be done, but  
is likely to have low sensitivity to certain models, since it requires
significant $|\ptmiss|$ as a trigger requirement. This would not
appear to be a {\em sine qua non}, however: at least in cases
where only one hadronically-decaying $\tau$ is present, one can
exploit the other lepton for the trigger.

\section{Summary}
We have argued that LHC searches for same-sign di-leptons of all
flavours provide a generic means by which one may discover 
supersymmetric scenarios without $R$-parity, which retain
naturalness of the weak scale (by keeping gluinos and third-generation
squarks light), but evade existing collider and flavour constraints.
The constraints are avoided by a combination of two factors: (i) a
reduction of $|\ptmiss|$ in final states because of
$R$-parity violating couplings; and (ii) heavy first- and
second-generation quarks.
The same-sign di-leptons are either provided by same-sign tops in decay chains, 
by lepton number violating couplings or by charged Higgs or $W$ decays in
decay chains. 

We have shown that same sign di-lepton searches provide essentially guaranteed
coverage of all but one scenario, in which the gluino decays
predominantly into a left-handed bottom squark that decays via the
charged Higgs into a virtual right-handed top squark, which in turn
decays to down-type quark jets via  the $U_3D_iD_j$ superpotential
operator(s). 
This exceptional case can only be missed 
in the small (and unlikely) part of parameter space where $\tan \beta
\ll 3$ and the charged Higgs
has no significant branching ratio for decay into either tau or top. 

We have assessed the impact of existing searches for same-sign di-leptons, designed for other
scenarios of new physics, in the case where pair-produced gluinos
decay to light, right-handed stop quarks, which in turn decay to
down-type quark jets via $U_3D_iD_j$ operators. The most stringent search
presents a bound of 
550 GeV on the gluino mass, approximately independently of the stop
mass. This approximation ought to be precise to a few tens of GeV, but some
of the less stringent bounds are 
crude estimates, since then the detector simulation is crude.
In some cases, we 
interpret the results of searches which have been
applied to very different models of new physics, and systematic errors on the
signal will be  
different to those 
assumed in the paper. 
Such changes in these signal systematic errors are neglected in our analysis.

We hope that our arguments regarding the importance of such
supersymmetric models are
sufficiently persuasive, and the results of our simulations
sufficiently promising, as to convince the experimental collaborations
to re-interpret the results of their existing same-sign di-lepton
searches in the context of the models described here. 
The recently developed {\tt RECAST} framework \cite{Cranmer:2010hk}
would seem to be an ideal vehicle by which to facilitate such
re-interpretations. 

Existing searches were designed with rather different models of
new physics in mind and it is clear that increased sensitivity could
be obtained by optimising them for the models that we suggest. In particular a
number of directions suggest themselves, namely by varying cuts on
$|\ptmiss|$, number of $b$ tags, and jet activity. Finally, we have
shown that a natural expectation in some scenarios is an excess of same-sign
$\tau$ final states over SM backgrounds. These may occur in a
sizable fraction of signal events 
(half of all events, for example, in decays via $Q_iL_3D_j$ operators),
which will more than offset the inevitable reduction in efficiency in
this channel compared to that for pairs of $e$ or $\mu$. A priority
for searches in this channel is the implementation of a $\tau$ (or
di-$\tau$) trigger which does not require large $|\ptmiss|$.
\section*{Acknowledgements}
This work has been partially supported by STFC\@. We would like to thank other
members of the Cambridge SUSY Working Group, J.~Butterworth, C.~Campagnari,
and R.~Sundrum for useful discussions.  

\bibliographystyle{JHEP-2}
\bibliography{a}

\end{document}